\documentclass[12pt]{article}
\usepackage{amsmath,amssymb,theorem,cite,epsfig,url,psfrag,eepic,amsmath,mathtools}
\advance \topmargin by -\headheight
\advance \topmargin by -\headsep
\evensidemargin \oddsidemargin
\def\Maketitle{{\def\newpage{}\maketitle}}
\makeatletter
\def\Appendix{\appendix
  \def\@seccntformat##1{Appendix~\csname the##1\endcsname.~~}}
\makeatother
\makeatletter
\@addtoreset{equation}{section}

\makeatother

\def\XXint#1#2#3{{\setbox0=\hbox{$#1{#2#3}{\int}$}
\vcenter{\hbox{$#2#3$}}\kern-.5\wd0}}

\begin{document}
\vspace*{0.3cm}
\vspace*{0.7cm}
\rightline{Dedicated to the memory of Lev Lipatov}
\title{\textbf{Integrability, Duality and Sigma Models}\vspace*{.3cm}}
\author{V.~A.~Fateev$^{1,2}$ and A.~V.~Litvinov$^{2,3}$\\[\medskipamount]
\parbox[t]{0.85\textwidth}{\normalsize\it\centerline{1. Laboratoire Charles Coulomb, Universit\'e de Montpellier, 34095  Montpellier, France}}\\
\parbox[t]{0.85\textwidth}{\normalsize\it\centerline{2. Landau Institute for Theoretical Physics, 142432 Chernogolovka, Russia}}\\
\parbox[t]{0.85\textwidth}{\normalsize\it\centerline{3. National Research University Higher School of Economics,119048 Moscow, Russia}}
}
\date{}
\Maketitle
\abstract{We introduce and study conformal field theories specified by $W-$algebras commuting with certain set of screening charges. These CFT's possess perturbations which define integrable QFT's. We establish that these QFT's have local and non-local Integrals of Motion and admit the perturbation theory in the weak coupling region. We construct factorized scattering theory which is consistent with non-local Integrals of Motion and perturbation theory. In the strong coupling limit the $S-$matrix of this  QFT tends to the scattering matrix of the $O(N)$ sigma model. The perturbation theory, Bethe anzatz technique, renormalization group approach and methods of conformal field theory are applied to show, that the constructed QFT's are dual to integrable deformation of $O(N)$ sigma-models.}
\section{Introduction}
Duality is an important concept of modern quantum field theory. Examples of the dualities  include: Kramers-Wanier duality of 2D Ising model, duality between sine-Gordon and Thirring models (bosonization), electric-magnetic duality in supersymmetric gauge theories, AdS/CFT correspondence etc. Known for many years, this phenomenon still looks rather mysterious and deserves further studies.  The models which admit dual strong coupling description can be studied in a wider range of the coupling constant region.  The phenomenon of  integrability plays an important role in justification of the dualities. In this paper we introduce and study  the duality between two class of  integrable quantum field theories. The former corresponds to integrable deformation of the $O(N)$ non-linear sigma model, while the other to integrable  QFT with Toda like interaction. For $N=3$ this duality relates the so called sausage sigma-model  \cite{Fateev:1992tk} with special integrable perturbation of sine-Liouville conformal field theory (see \cite{Fateev:2017mug} for review of this CFT and its integrable perturbations). The $N=4$ cousin of this duality has been introduced and studied  in  \cite{Fateev:1995ht,Fateev:1996ea}. Here we provide a generalization for $N>4$. 

One can explain the origin of the duality studied here using the $S$-matrix formalism. Integrability in two-dimensional QFT puts strong constraints on scattering. In particular, it implies the property of factorization of multi-particle processes into the product of two-particle ones. The two-particle amplitudes are further constrained by integrability, global symmetries and  conditions of unitarity and crossing. In particular, the $S$-matrix is forced to obey the Yang-Baxter equation. There are so called trigonometric solutions to the YB equation, which depend on a continuos parameter $\lambda$. At some value of this parameter $\lambda=\lambda_{\textrm{free}}$ the $S$-matrix becomes identical and it may exists a Lagrangian theory which describes the behavior near this free point. But it can be also another point $\lambda=\lambda_{\textrm{rat}}$, where the $S$-matrix becomes the rational one. Usually, rational $S$-matrices with properly chosen CDD factors correspond to sigma-models on symmetric spaces. Near the point $\lambda=\lambda_{\textrm{rat}}$ the theory can be considered as a deformation of this sigma-model. 

In these notes we consider trigonometric solutions to the YB equation related  to the quantum affine group $U_{q}(\widehat{\mathfrak{so}}(N))$. Corresponding scattering matrix has both limits mentioned above.  In one limit  it tends to the rational one and coincides with the exact $S-$matrix for the $O(N)$ sigma-model \cite{Zamolodchikov:1978xm}. On the other hand the same $S$-matrix degenerates to the identity  matrix in another limit. The main result of this paper is the construction of the  Lagrangian theory whose perturbative expansion describes the vicinity of this free point.

As we will see below, this later theory can be interpreted as a perturbed CFT.  The corresponding CFT is governed by a certain $\mathbf{W}$-algebra. Its universal enveloping algebra has a Cartan subalgebra which consists of an infinite number of local Integrals of Motion of odd spins.  This integrable system is inherited in perturbed theory. In particular, it opens the possibility to study the finite volume spectrum using the conformal perturbation theory. From the other hand this CFT has a Lagrangian description which can be used to compute important characteristics of CFT called the reflection amplitudes.

The deformed sigma model can be studied within RG approach \cite{Friedan:1980jm}. One can use this approach to compare with the data obtained from the Bethe Anzatz method from scattering theory. The agreement of both methods is another justification of the duality. One can study the short distance behavior within this approach and compute classical version of reflection amplitudes and compare them to the limit of exact expressions.

This paper organized as follows. In section \ref{CFT} we introduce and study CFT of Toda type, which has hidden Lie algebraic structure. In particular, using free-field representation we compute reflection amplitudes in this theory. In section \ref{dual-L} we introduce QFT which can be treated as certain integrable perturbation of CFT studied in section \ref{CFT}. We claim that this theory describes scattering of particles with $U_{q}(\widehat{\mathfrak{so}}(N))$ trigonometric $S$-matrix and hence it can be considered as  a dual theory for the deformed $O(N)$ sigma-model. In section \ref{TBA} we make this relation more precise and derive the exact relation between the parameters of the Lagrangian and $S-$matrix. In section \ref{Ricci-flow} we study sigma-models using RG approach and present solutions to Ricci flow equation corresponding to the deformed sigma-models. In section \ref{MSS} we perform computations in the so called minisuperspace approximation to confirm the identification of the deformed sigma-models with factorized scattering theory.
\section{Conformal field theory, reflection amplitudes}\label{CFT} 
In this section we consider CFT which has hidden Lie algebra symmetry and can be described by $2m$ massless bosonic fields
\begin{equation*}
  \{\varphi,\phi\}=\{(\varphi_{1},\dots,\varphi_{m}),(\phi_{1},\dots,\phi_{m})\}.
\end{equation*}
and parameters $a$ and $b$, satisfying the condition
\begin{equation}
  a^{2}-b^{2}=1.
\end{equation}
We introduce two $m-$dimensional vectors $\rho$ and $\rho_{1}$, which are the halves of sum of the roots of $b(m)$ and $d(m)$
\begin{equation}\label{ro}
  \rho=\left(\frac{1}{2},\frac{3}{2},\dots,\frac{2m-1}{2}\right),\quad
  \rho_{1}=\left( 0,1,2,\dots,m-1\right)
\end{equation}
and the dilaton field
\begin{equation*}
  \Phi_{d}=\frac{1}{b}(\rho\cdot\varphi)+\frac{i}{a}(\rho_{1}\cdot \phi)=(\mathrm{Q}\cdot\varphi)+i(\mathrm{Q}_{1}\cdot\phi).
\end{equation*}
Then $m$ pairs of ``screening fields'' $\{\mathrm{V}_{1,1,}\mathrm{V}_{1,-1}\},\{\mathrm{V}_{-1,2,}\mathrm{V}_{2,-2}\},\dots,\{\mathrm{V}_{-(m-1),m},\mathrm{V}_{m,-m}\}$:
\begin{equation}\label{scr}
\begin{gathered}
  \{\mu\exp(b\varphi_{1}+ia\phi_{1}),\mu\exp(b\varphi_{1}-ia\phi_{1})\},\,\{\mu\exp(-b\varphi_{1}+ia\phi_{2}),\mu\exp(b\varphi_{2}-ia\phi_{2})\},\\
    \dots,\{\mu\exp(-b\varphi_{m-1}+ia\phi_{m}),\mu\exp(b\varphi_{m}-ia\phi_{m})\}
\end{gathered}
\end{equation}
define  CFT with the action
\begin{equation}\label{ac2}
  \mathcal{A}_{CFT}=\int\left(\frac{1}{8\pi}\Bigl((\partial_{\mu}\varphi\cdot\partial_{\mu}\varphi)+
  (\partial_{\mu}\phi\cdot\partial_{\mu}\phi)\Bigr)+\mathrm{U}^{(m)}(\varphi,\phi)+R_{2}\Phi_{d}\right)d^{2}x,
\end{equation}
where $\mathrm{U}^{(m)}(\varphi,\phi)=\mathrm{V}_{1,1}+\mathrm{V}_{1,-1}+\mathrm{V}_{-1,2}+\mathrm{V}_{2,-2}+\dots+\mathrm{V}_{-(m-1),m}+\mathrm{V}_{m,-m}$, and $R_{2}$ is the two dimensional scalar curvature of the world sheet. The central charge of this CFT is
\begin{equation}\label{centra-charge}
c=m\left(2+\frac{(2m-1)(2m+1)}{2b^{2}}-\frac{(m-1)(2m-1)}{a^{2}}\right)
\end{equation}
and the primary fields are the exponential fields
\begin{equation*}
  \mathrm{V}_{\mathrm{A,B}}=\exp\left((\mathrm{A}\cdot\varphi)+i(\mathrm{B}\cdot\phi)\right)
\end{equation*}
with the conformal dimensions
\begin{equation*}
   \Delta(\mathrm{A,B})=\frac{1}{2}\left(-\mathrm{A}^{2}+2(\mathrm{A}\cdot\mathrm{Q})+\mathrm{B}^{2}-2(\mathrm{B}\cdot\mathrm{Q}_{1})\right).
\end{equation*}
These fields are normalized by the condition
\begin{equation*}
  \langle\mathrm{V}_{\mathrm{A,B}}(x)\mathrm{V}_{2\mathrm{Q}-\mathrm{A},2\mathrm{Q}_{1}-\mathrm{B}}(0)\rangle=|x|^{-4\Delta(\mathrm{A},\mathrm{B})}.
\end{equation*}

In the space of the fields $\mathrm{V}_{\mathrm{A},\mathrm{B}}$ one can define the action of  the Weil group $\mathrm{w}_{m}$ of the Lie algebras $b(m)(c(m))$. It acts as $\mathrm{V}_{\mathrm{A},\mathrm{B}}\rightarrow\mathrm{R}_{\widehat{s}}\mathrm{V}_{\mathrm{Q}+\widehat{s}(\mathrm{A}-\mathrm{Q),\mathrm{B}}}$, where $\mathrm{R}_{\widehat{s}}$ is the reflection
amplitude. The reflection amplitudes are the identification carts of CFTs, admitting bosonization. Two point correlation function which coincides with the ``maximal'' reflection amplitude
($\mathrm{a}\rightarrow-\mathrm{a}$):
\begin{equation*}
    \langle\mathrm{V}_{\mathrm{Q+a,Q}_{1}\mathrm{+b}}(x)\mathrm{V}_{\mathrm{Q+a,\mathrm{Q}_{1}-b}}(0)\rangle=\mathrm{R(a,b)}|x|^{-4\Delta}%
\end{equation*}
in the CFT \eqref{ac2} can be calculated using the integral relation \cite{Baseilhac:1998eq,Fateev:2007qn}
\begin{multline}\label{Fateev-integral}
 \int\mathcal{D}_{n}(x)\prod_{i=1}^{n}\prod_{j=1}^{n+m+2}|x_{i}-t_{j}|^{2p_{j}}\,d^{2}\vec{x}_{n}=
 \prod_{j=1}^{n+m+2}\gamma(1+p_{j})\prod_{i<j}|t_{i}-t_{j}|^{2+2p_{i}+2p_{j}}
 \times\\\times
 \int\mathcal{D}_{m}(y)\prod_{i=1}^{m}\prod_{j=1}^{n+m+2}|y_{i}-t_{j}|^{-2-2p_{j}}\,d^{2}\vec{y}_{m},
\end{multline}
where
\begin{equation*}
  \mathcal{D}_{n}(x)=\prod_{i<j}|x_{i}-x_{j}|^{2},\qquad
  d^{2}\vec{x}_{n}=\frac{1}{\pi^{n}n!}\prod_{j=1}^{n}d^{2}x_{j},\quad
  \gamma(x)=\frac{\Gamma(x)}{\Gamma(1-x)}\quad\text{and}\quad
  \sum_{j=1}^{n+m+2} p_{j}=-n-1.
\end{equation*}

To describe it we introduce the notations: $e_{\alpha}$ are the positive roots of $c(m)$, $\widehat{e}_{\alpha}=2e_{\alpha}/(e_{\alpha}\cdot e_{\alpha})$ are the dual roots, $e_{j}$ the simple roots, and the vectors $\omega_{i}$ satisfying the condition $(\omega_{i}\cdot e_{j})=\delta_{j}^{i}$. Then the function $\mathrm{R(a,b)}$ can be represented in the form
\begin{equation}\label{ra}
 \mathrm{R(a,b)=}\frac{\mathrm{A(-a,b)}}{\mathrm{A(a,b)}},
\end{equation}
where
\begin{equation}\label{ra1}
    \mathrm{A(a,b)}=\prod\limits_{j=1}^{m}\left(\frac{2\pi\mu}{(e_{j}\cdot e_{j})b^{2}}\right)^{-(\mathrm{a}\cdot\mathrm{\omega}_{j})/b}
    \prod\limits_{i=1}^{j}\frac{\Gamma\left(  \frac{1}{2}+b\mathrm{a}_{j}+a\mathrm{b}_{i}\right)}{\Gamma\left(  \frac{1}{2}-\mathrm{a}_{j}+a\mathrm{b}_{i}\right)}
    \prod\limits_{\alpha>0}\Gamma(-b(\mathrm{a}\cdot e_{\alpha}))\Gamma(-\frac{1}{b}(\mathrm{a}\cdot\widehat{e}_{\alpha}))
\end{equation}
The general reflection amplitude can be written as
\begin{equation}\label{RA}
 \mathrm{R}_{\widehat{s}}\mathrm{(a,b)=}\frac{\mathrm{A}(\widehat{s}(\mathrm{a}),\mathrm{b)}}{\mathrm{A(a,b)}}.
\end{equation}
We note that the reflection amplitudes are not only the ``visit carts'' of $W-$symmetries of the CFTs, but also play an important role for the calculation of vacuum expectation values of the exponential fields and UV asymptotics in the perturbed CFTs.

At the end of this section we note that the symmetry algebra of  CFT considered above can be extended by the introduction of additional field $\phi_{m+1}$, and the
additional term to the action \eqref{ac2}
\begin{equation}\label{ac2-2}
  \mathcal{A}_{CFT}^{\prime}=\mathcal{A}_{CFT}+\int\left(\frac{1}{8\pi}(\partial_{\mu}\phi_{m+1}\cdot\partial_{\mu}\phi_{m+1})\right)d^{2}x.
\end{equation}
In next section we will consider integrable perturbed QFT's which can be viewed as perturbed CFT's \eqref{ac2} and \eqref{ac2-2}.
\section{Integrable perturbed CFT}\label{dual-L}
In this section we consider QFTs, which can be derived from CFTs \eqref{ac2}, \eqref{ac2-2}  by neglecting the dilaton and adding the affine terms. Namely we denote 
\begin{equation}\label{perturbation}
  \begin{aligned}
    &\mathbf{U}_{odd}^{(m)}\left(\varphi,\phi\right)=\mathrm{U}^{(m)}\left(\varphi,\phi\right)+\mu\exp(-b\varphi_{m}-ia\phi_{m}),\\
    &\mathbf{U}_{ev}^{(m)}\left(\varphi,\phi\right)=\mathrm{U}^{(m)}\left(\varphi,\phi\right)+\mu\bigl(\exp(-b\varphi_{m}+ia\phi_{m+1})+\exp(-b\varphi_{m}-ia\phi_{m+1})\bigr).
  \end{aligned}
\end{equation}
The notations are related with the property that $\mathbf{U}_{odd}^{(m)}\left(\varphi,\phi\right)$ and $\mathbf{U}_{ev}^{(m)}\left(\varphi,\phi\right)$ correspond to QFT's dual to the deformed $O(n+2)-$models with odd $n=2m-1$ and even $n=2m$.  As the non-local integrals of motion and $S-$matrices for these cases are slightly different we consider them separately. We give examples of local Integrals of Motion in appendix \ref{IM}.
\subsection{The case $n=2m-1$}
In the case $n=1$, the potential $\mathbf{U}^{(1)}_{odd}(\varphi,\phi)$ is special. It has the form
\begin{equation*}
    \mathbf{U}^{(1)}_{odd}(\varphi,\phi)=2\mu\exp(b\varphi)\cos(a\phi)+2\mu\exp(-b\varphi)\cos(a\phi)
\end{equation*}
The QFT with the action
\begin{equation}\label{A1}
   \mathcal{A}=\int\left(\frac{1}{8\pi}\bigl((\partial_{\mu}\varphi\cdot\partial_{\mu}\varphi)+(\partial_{\mu}\phi\cdot\partial_{\mu}\phi)\bigr)+
   \mathbf{U}^{(1)}_{odd}\left(\varphi,\phi\right)\right)d^{2}x.
\end{equation}
is known to be dual to the deformed $O(3)$ sigma-model, or the sausage sigma-model. The scattering theory, TBA analysis and the metric of the corresponding deformed $O(3)$-sigma model can be found in \cite{Fateev:1992tk,Fateev:1995ht}. Besides the local Integrals of Motion this QFT has the non-local ones generated by the non-local fields 
\begin{equation*}
 J_{\pm}=e^{\pm i\hat{\phi}/a}(b\partial\varphi+ia\partial\phi);\qquad I_{\pm}=e^{\pm\hat{\varphi}/b}(b\partial\varphi+ia\partial\phi)
\end{equation*}
where $\hat{\varphi}$ and $\hat{\phi}$ are the chiral parts of fields $\varphi$ and $\phi$. To develop the perturbation theory for small $b$ with the action \eqref{A1} it is convenient to use 2D Mandelstam-Coleman correspondence \cite{Coleman:1974bu,Mandelstam:1975hb} between bosons and fermions
\begin{equation}\label{cm}
  \frac{1}{2}\partial_{\mu}\phi\partial_{\mu}\phi\rightarrow i\bar{\psi}\gamma_{\mu}\partial_{\mu}\psi;\quad
  \partial_{\mu}\phi\rightarrow i\varepsilon_{\mu\nu}\bar{\psi}\gamma_{\nu}\psi,\quad 
  e^{\pm i\phi}=\bar{\psi}(1\pm\gamma_{5})\psi,
\end{equation}
and rewrite \eqref{A1} in the form
\begin{subequations}
\begin{equation}
   \mathcal{A}=\int d^{2}x\left(L_{F}+L_{FB}+L_{B}\right),
\end{equation}
where
\begin{equation}
\begin{gathered}
  L_{F} =\frac{1}{4\pi}\left(i\bar{\psi}\gamma_{\mu}\partial_{\mu}\psi-\frac{b^{2}}{8(1+b^{2})}(\bar{\psi}\gamma_{\nu}\psi)^{2}\right);\quad 
  L_{FB}=\frac{M_{0}}{4\pi}\bar{\psi}\psi\cosh(b\varphi);\\
  L_{B}=\frac{1}{8\pi}\left((\partial_{\mu}\varphi,\partial_{\mu}\varphi)+\frac{M_{0}^{2}}{b^{2}}\sinh^{2}(b\varphi)\right),\qquad
  \frac{M_{0}}{4\pi}=\mu.
\end{gathered}
\end{equation}
\end{subequations}
Here the last term in $L_{B}$ is the usual counterterm, which cancels the divergencies coming from the fermion loops.

For odd $n=2m-1>1$ the theory with $m$ ``compact'' fields $\phi$ and $m$ ``non-compact'' fields $\varphi$ is described by the action
\begin{equation}\label{ann}
  \mathcal{A}=\int\left(\frac{1}{8\pi}\bigl((\partial_{\mu}\varphi\cdot\partial_{\mu}\varphi)+(\partial_{\mu}\phi\cdot\partial_{\mu}\phi)\bigr)+
  \mathbf{U}_{odd}^{(m)}(\varphi,\phi)\right)d^{2}x.
\end{equation}
It possesses the non-local integrals of motions generated by the fields
\begin{equation}\label{intm}
\begin{gathered}
  J_{0}=e^{-i(-\hat{\phi}_{1}-\hat{\phi}_{2})/a}\left(b\partial\varphi_{1}+ia\partial\phi_{1}\right),\quad 
  J_{i}=e^{-i(\hat{\phi}_{i}-\hat{\phi}_{i+1})/a}\left(  b\partial\varphi_{i}-ia\partial\phi_{i+1}\right),\quad i=1,..,m-1,\\
  J_{m}=e^{-i\hat{\phi}_{m}/a}(b\partial\varphi_{m}+ia\partial\phi_{m})
\end{gathered}
\end{equation}
and
\begin{equation}
\begin{gathered}
    I_{0}=e^{\hat{\varphi}_{1}/b}\left(  b\partial\varphi_{1}+ia\partial\phi_{1}\right),\quad 
    I_{i}=e^{(\hat{\varphi}_{i+1}-\hat{\varphi}_{i})/b}\left(b\partial\varphi_{i+1}-ia\partial\phi_{i+1}\right),\quad i=1,..,m-1,\\ 
    I_{m}=e^{(\hat{\varphi}_{_{m}}+\hat{\varphi}_{m-1})/b}\left(  b\partial\varphi_{_{_{m}}}+ia\partial\phi_{m}\right).
\end{gathered}
\end{equation}
We note that the Integrals of motion $\mathcal{J}_{s}$ constructed from the fields $J_{s}$  have the spins $\Delta_{s}=\frac{1}{a^{2}}$, $s=0,1,\dots,\frac{n-1}{2}$,
and $\Delta_{m}=\frac{1}{2a^{2}}$. The integrals $\mathcal{J}_{1},\dots,\mathcal{J}_{m}$ form the Borel subalgebra of the quantum group $b\left(m\right)_{q}$, with $q=e^{i2\pi/a^{2}}$. The integral $\mathcal{J}_{0}$ adds the affine generator. It looks very natural that QFT (\ref{ann}) possesses affine $b\left(m\right)  _{q}$ symmetry. 

It is useful to rewrite the action \eqref{ann} in the form convenient for the perturbation theory, using the boson-fermion correspondence \eqref{cm}. We denote $\gamma_{\pm}=\frac{1}{2}(1\pm\gamma_{5})$, then
\begin{subequations}\label{Fac}
\begin{equation}
  \mathcal{A}=\int\left(  L_{F}+L_{FB}+L_{B}\right)d^{2}x,
\end{equation}
where
\begin{equation}
\begin{gathered}
   L_{F}=\frac{1}{4\pi}\sum_{k=1}^{m}\left(  i\bar{\psi}_{k}\gamma_{\mu}\partial_{\mu}\psi_{k}-\frac{b^{2}}{8(1+b^{2})}(\bar{\psi}_{k}\gamma_{\nu}\psi_{k})^{2}\right),\\
   L_{FB}=\frac{M_{0}}{4\pi}\left(  e^{b\varphi_{1}}\bar{\psi}_{1}\psi_{1}+
   \sum\limits_{i=2}^{m}\bigl(e^{-b\varphi_{i-1}}\bar{\psi}_{i}\gamma_{+}\psi_{i}+e^{b\varphi_{i}}\bar{\psi}_{i}\gamma_{-}\psi_{i}\bigr)
   +e^{-b\varphi_{m}}\bar{\psi}_{_{m}}\gamma_{-}\psi_{_{m}}\right)\\
   L_{B}=\frac{1}{8\pi}\left(  (\partial_{\mu}\varphi\cdot\partial_{\mu}\varphi)+\frac{M_{0}^{2}}{b^{2}}\left(  e^{2b\varphi_{1}}+
   2\sum\limits_{i=2}^{m-1}e^{b(\varphi_{i}-\varphi_{i-1})}+e^{b(\varphi_{m}-\varphi_{m-1})}+e^{b(\varphi_{m}+\varphi_{m-1})}\right)\right).
\end{gathered}
\end{equation}
\end{subequations}
The potential term in this equation is the usual counerterm, which cancels the divergencies coming from fermion loops. The term $L_{B}$ is the Toda theory with affine the Lie algebra dual to affine $b(m)$, which we call $\hat{b}(m)$. The spectrum of this Toda theory for small $b^{2}$ is $M_{a}=2M_{0}\sin\left(\frac{\pi(a-1)}{n}\right)+O(b^{2})$, $2\leq a<m+1,\quad M_{m+1}=M_{0}+O(b^{2})$. The last particle and $2m$ fermionic particles $\psi_{i}$, $\psi_{i}^{\ast}$, $i=1,\dots,m$ form the $2m+1=n+2$ multiplet which permits us to construct the scattering theory with the affine $b(m)_{q}$ symmetry.

The scattering theory for QFT with the action \eqref{Fac} contains $n+2$ particles of the same mass $M$, which we denote $\{A_{i}(\theta)\}$ and $m-1$ ``bound states'' $M_{a}$ which disappear from the spectrum (as we will see later) for finite $b^{2}$. The parameter $\theta$ in the scattering amplitudes is the difference of the rapidities of the colliding particles. The theory \eqref{Fac} has the $U(1)^{m}$ symmetry $\psi_{i}\rightarrow e^{i\eta_{i}}\psi_{i}$, $\psi_{i}^{\ast}\rightarrow e^{-i\eta_{i}}\psi^{\ast}$ and the $\mathbf{C}$ symmetry $\psi_{i}\rightarrow\psi_{i}^{\ast}$ (and hence the $\mathbf{PT}$-symmetry). It is convenient to numerate the particles $A_{i}=\psi_{i}$, $i=1,\dots,m,$ $A_{n+3-i}=A_{\bar{\imath}}=\bar{A}_{i}=\psi_{i}^{\ast}$ and $A_{m+1}=M_{m+1}$.

We denote as $\bar{j}$ $=n+3-j$. The particles $\{A_{j},A_{\bar{j}}\}$, form $U(1)$ doublets, while the particle $A_{m+1}$ is neutral. Factorized $S-$matrix with the affine $b(m)_{q}$ symmetry and with $\mathbf{C}$, $\mathbf{PT}$ and crossing invariances has the form
\begin{equation}\label{un}
   S_{j_{1},j_{2}}^{i_{1,}i_{2}}(\theta)=S_{i_{2},i_{1}}^{j_{2,}j_{1}}(\theta),\quad S_{j_{1},j_{2}}^{i_{1,}i_{2}}(\theta)=S_{\bar{j}_{1},\bar{j}_{2}}^{\bar{i}_{1},\bar{i}_{2}}(\theta),\quad
  S_{j_{1},j_{2}}^{i_{1,}i_{2}}(\theta)=S_{i_{1},\bar{j}_{2}}^{j_{1},\bar{i}_{2}}\left(i\pi-\theta\right)=S_{j_{2}\bar{i}_{1}}^{i_{2}\bar{j}_{1}}\left(i\pi-\theta\right),
\end{equation}
and can be derived from the results of V.Bazhanov paper \cite{Bazhanov:1984gu}. This $S$-matrix depends on a continuos parameter $\lambda$. It can be expressed in term of the unitarizing factor $F\left(  \theta\right)  $ satisfying the conditions
\begin{equation}\label{unc}
   F(\theta)F(-\theta)  =1,\ F(i\pi-\theta)=F(\theta)\frac{\sinh\left(\frac{n}{2}\lambda\theta\right)\sinh\left(\frac{n}{2}\lambda(\theta-\frac{i\pi\left(n-2\right)\pi}{n})\right)}
   {\sinh\left(\frac{n}{2}\lambda(\theta-\frac{i2\pi}{n}\right)\sinh\left(\frac{n}{2}\lambda(\theta-i\pi)\right)},
\end{equation}
and the functions
\begin{equation}\label{trR}
  t(\theta)=\frac{\sinh\left(\frac{n}{2}\lambda\theta\right)}{\sinh\left(  \frac{n}{2}\lambda(\theta-\frac{i2\pi}{n}\right)},\, 
  r(\theta)=\frac{-i\sin\left(\pi\lambda\right)}{\sinh\left(\frac{n}{2}\lambda(\theta-\frac{2i\pi}{n})\right)},\,
  R(\theta)=\frac{-\sin\left(\pi\lambda\right)\sin\left(\pi\lambda\frac{n}{2}\right)}{\sinh\left(  \frac{n}{2}\lambda(\theta-\frac{2i\pi}{n})\right) \sinh\left(  \frac{n}{2}\lambda(\theta-i\pi)\right)}. 
\end{equation}
Namely up to the symmetries \eqref{un} the $S-$matrix has the following amplitudes
\begin{align}\label{Sb}
   S_{ii}^{ii}&=F\left(\theta\right),i\neq m+1;\ S_{m+1m+1}^{m+1m+1}=F(\theta)\left(t\left(\theta\right)+R\left(\theta\right)\right);\quad\text{for}\quad i\neq j,i\neq\bar{j}:\nonumber\\
   S_{ij}^{ij}  &  =F\left(\theta\right)t\left(\theta\right);\ S_{j\bar{j}}^{j\bar{j}}=F(i\pi-\theta);S_{ji}^{ij}=F\left(\theta\right)r\left(\theta\right)e^{i\kappa_{ij}\lambda\theta};\ 
   S_{j\bar{j}}^{i\bar{i}}=S_{ij}^{ji}(i\pi-\theta);\nonumber\\
   S_{\bar{1}1}^{1\bar{1}}& =S_{1\bar{1}}^{\bar{1}1}=F\left(  \theta\right)R\left(\theta\right);\, S_{\bar{i}i}^{i\bar{i}}=S\left(\theta\right)_{i-1,\bar{i}}^{\bar{i}i-1}+
   S_{\bar{i},i-1}^{i-1\bar{i}}\left(i\pi-\theta\right),i<\bar{i}.
\end{align}
The function $\kappa_{ij}=\kappa_{\bar{i}\bar{j}}=-\kappa_{ji}=-\kappa_{\bar{i}j}$, $\kappa_{1\bar{1}}=\kappa_{1m+1}=0$ is defined as:
$\kappa_{ij}=i-j-\frac{n}{2}\textrm{sgn}(i-j)$, $i,j<m+1$; $\kappa_{i,m+1}=m+1-i$.

The solution of unitarity and crossing symmetry equations \eqref{unc} in terms of the parameter $p=\frac{1}{\lambda}$ has the form
\begin{equation}\label{F}
  -F\left(\theta\right)=\exp\left(  i\delta_{n,p}\left(  \theta\right)\right)=\exp\left(i\int\limits_{-\infty}^{\infty}\frac{d\omega\cosh\left(\frac{\pi\omega(n-2)}{2n}\right)  
  \sinh\left(\frac{\pi\omega(p-1)}{n}\right)  \sin\omega\theta}{\omega\cosh\left(\frac{\pi\omega}{2}\right)  \sinh\left(  \frac{\pi\omega p}{n}\right)}\right)
\end{equation}
The comparison with perturbation theory gives $p=\frac{1}{\lambda}=1+b^{2}+\dots$ In the limit $p\rightarrow\infty$ the scattering matrix tends to the $S-$matrix of the $O(n+2)-$sigma model.
\subsection{The case of $n-$even, $n=2m$}
For even $n=2m$ the theory with $m+1$ ``compact'' fields $\phi$ and $m$ ``non-compact'' fields $\varphi$ can be described by the action
\begin{equation}\label{acd}
  \mathcal{A}=\int\left(\frac{1}{8\pi}\bigl((\partial_{\mu}\varphi\cdot\partial_{\mu}\varphi)+(\partial_{\mu}\phi\cdot\partial_{\mu}\phi)\bigr)+
  \mathbf{U}_{ev}^{(m)}(\varphi,\phi)\right)d^{2}x.
\end{equation}
It has the non-local integrals of motions generated by the fields
\begin{equation}\label{Id}
\begin{gathered}
  J_{0} =e^{-i(-\hat{\phi}_{1}-\hat{\phi}_{2})/a}\left(b\partial\varphi_{1}+ia\partial\phi_{1}\right),\quad 
  J_{i}=e^{-i(\hat{\phi}_{i}-\hat{\phi}_{i+1})/a}\left(b\partial\varphi_{i}-ia\partial\phi_{i+1}\right),\,i=1,\dots,m,\\
  J_{m+1}=e^{-i(\hat{\phi}_{m}+\hat{\phi}_{m+1})/a}(b\partial\varphi_{m}+ia\partial\phi_{m+1}),
\end{gathered}
\end{equation}
and
\begin{equation}
\begin{gathered}
   I_{0}=e^{\hat{\varphi}_{1}/b}\left(  b\partial\varphi_{1}+ia\partial\phi_{1}\right),\quad 
   I_{i}=e^{(\hat{\varphi}_{i+1}-\hat{\varphi}_{i})/b}\left(  b\partial\varphi_{i+1}-ia\partial\phi_{i+1}\right),\,i=1,\dots,m,\\ 
   I_{m}=e^{\hat{\varphi}_{m}/b}\left(  b\partial\varphi_{m}+ia\partial\phi_{m+1}\right).
\end{gathered}
\end{equation}
The Integrals of motion $\mathcal{J}_{s}$ constructed with fields $J_{s}$ have the spins $\Delta_{s}=\frac{1}{a^{2}}$ ,$ s=0,1,\dots,m+1$. The integrals $\mathcal{J}_{1},\dots,\mathcal{J}_{m+1}$ form the Borel subalgebra of the quantum group $d(m+1)_{q}$ with $q=e^{i2\pi/a^{2}}$. The integral $\mathcal{J}_{0}$ adds the affine generator. It means that QFT \eqref{acd} possesses affine $d(m+1)_{q}$ symmetry. We use Eq. \eqref{cm} to rewrite the action \eqref{acd} in the form convenient for the perturbation theory
\begin{subequations}\label{afd}
\begin{equation}
\mathcal{A}=\int\left(L_{F}+L_{FB}+L_{B}\right)d^{2}x
\end{equation}
where
\begin{equation}\label{afd-2}
\begin{gathered} 
  L_{F}=\frac{1}{4\pi}\sum_{k=1}^{m+1}\left(  i\bar{\psi}_{k}\gamma_{\mu}\partial_{\mu}\psi_{k}-\frac{b^{2}}{8(1+b^{2})}(\bar{\psi}_{k}\gamma_{\nu}\psi_{k})^{2}\right),\\
  L_{FB}=\frac{M_{0}}{4\pi}\left(  e^{b\varphi_{1}}\bar{\psi}_{1}\psi_{1}+
  \sum\limits_{i=2}^{m}(e^{-b\varphi_{i-1}}\bar{\psi}_{i}\gamma_{+}\psi_{i}+e^{b\varphi_{i}}\bar{\psi}_{i}\gamma_{-}\psi_{i})+e^{-b\varphi_{m}}\bar{\psi}_{m+1}\psi_{m+1}\right),\\
  L_{B}=\frac{1}{8\pi}\left((\partial_{\mu}\varphi\cdot\partial_{\mu}\varphi)+\frac{M_{0}^{2}}{b^{2}}\left(  e^{2b\varphi_{1}}+2\sum\limits_{i=2}^{m}e^{b(\varphi_{i}-\varphi_{i-1})}+
  e^{-2b\varphi_{m}}\right)\right).
\end{gathered}
\end{equation}
\end{subequations}
The last term in equation \eqref{afd-2} is the usual counterterm, which cancels the divergencies coming from fermion loops. The term $L_{B}$ is the Toda theory with the affine Lie algebra $c(m)$. The spectrum of this Toda theory for small $b^{2}$ is $M_{a}=2M_{0}\sin\left(\frac{\pi a}{n}\right)+O(b^{2})$, $1\leq a\leq m$. The
fermionic particles $\psi_{i},\psi_{i}^{\ast}$ $i=1,\dots,m+1$ form the $2m+2=n+2$ multiplet which permits us to construct the scattering theory with the affine $d(m+1)_{q}$ symmetry. We again denote $\bar{j}=n+3-j$ and numerate the particles $A_{i}=\psi_{i},i=1,\dots,m=1$, $A_{n+3-i}=A_{\bar{\imath}}=\psi_{i}^{\ast}$. The scattering theory consistent with the perturbation theory, with $U(1)^{m+1}$ and $\mathbf{C,PT}$ invariances and with the affine $d(m+1)_{q}$ symmetry is
\begin{align}\label{fst2}
  S_{ii}^{ii}&=F(\theta);\ S_{ij}^{ij}=F(\theta)t(\theta),i\neq\bar{j};\ S_{j\bar{j}}^{j\bar{j}}=F(i\pi-\theta);\quad for\quad i\neq j\neq\bar{j}\nonumber\\
  S_{ji}^{ij}&=F(\theta)r(\theta)e^{i\kappa_{ij}\lambda\theta};\ S_{j\bar{j}}^{i\bar{i}}=S_{ij}^{ji}(i\pi-\theta);S_{\bar{1}1}^{1\bar{1}}=S_{\overline{m+1}m+1}^{m+1\overline{m+1}}=
  F(\theta)R(\theta);\nonumber\\
  S_{\bar{1}1}^{1\bar{1}} &=S_{\overline{m+1}m+1}^{m+1\overline{m+1}}=F(\theta)R(\theta);\ S_{\bar{i}i}^{i\bar{i}}=S_{\bar{i}-1i-1}^{i-1\bar{i}-1}(\theta)+
  S_{i-1,\bar{i}-1}^{\bar{i}-1i-1}(i\pi-\theta),i\neq1,n.
\end{align}
The function $\kappa_{ij}=\kappa_{\bar{i}\bar{j}}=-\kappa_{ji}=-\kappa_{\bar{i}j}$, $\kappa_{1\bar{1}}=\kappa_{1m+1}=0$; $\kappa_{ij}=i-j-\frac{n}{2}\textrm{sgn}(i-j)$, 
$i,j<m+1$. The function $F(\theta)$ is given by the same expression \eqref{F}, where $p=\frac{1}{\lambda}$.
\section{Non-perturbative consideration}\label{TBA}
In this Section we consider the cases of odd and even $n$ in the same way. To derive the exact relations between the parameters of the action and of the scattering theory it is important to calculate the observables using both approaches. Our QFTs possess $U(1)^{[\frac{n+2}{2}]}$ symmetry generated by the charges $Q_{j}=\int\psi_{j}\psi_{j}^{\ast}dx$. One can add to the Hamiltonian of our QFT the terms $-A_{i}Q_{i}$ with different parameters $A_{i}$ (chemical potentials). The simplest way to calculate the ground state energy using BA approach is to consider the configurations of the chemical potentials $A_{i}$ which lead to condensation of  particles, which have the simple scattering amplitudes, namely the pure phases. In our scattering theories it corresponds the case when only one kind of particles is condensed. This configuration corresponds to $A_{1}=A$ and all other $A_{i}=0.$ When $A\gg M$ we can neglect in the actions \eqref{Fac} and \eqref{afd} all terms containing massive parameter $M_{0}$ and only the particle $\psi_{1}$ will condense. It means that for calculation of the density of ground state energy we can put all $\psi_{i}$ with $i>1$ equal to zero. The action in this case coincides with the action of massless Thirring model:
\begin{equation}
 \mathcal{A}_{TM}=\frac{1}{4\pi}\int\left(i\overline{\psi}_{1}\gamma_{\mu}\partial_{\mu}\psi_{1}-\frac{b^{2}}{8(1+b^{2})}(\overline{\psi}_{1}\gamma_{\mu}\psi_{1})^{2}\right)d^{2}x
\end{equation}
At $A\rightarrow\infty$ the ground state energy $\mathcal{E}(A)$ approaches the ground state energy of massless Thirring model  with the coupling constant $g=-\frac{b^{2}}{(1+b^{2})}$
\begin{equation}\label{ass1}
  \mathcal{E}(A)\underset{A\rightarrow\infty}{\longrightarrow}\mathcal{E}_{TM}=-\frac{A^{2}}{2\pi}(1+g)^{-1}=-\frac{(1+b^{2})A^{2}}{2\pi}.
\end{equation}
We calculate now the same quantity from the BA approach. For $A>M$ we have the sea of particles $\psi_{1}(\theta)$ which fill all possible states inside some ``Fermi interval'' $-B<\theta<B$. The distribution of particles $\epsilon(\theta)$ within this interval is determined by their scattering amplitude $S_{11}^{11}(\theta)=-F_{n}(\theta)$. The specific ground state energy has the form
\begin{equation}\label{en}
   \mathcal{E}(A)-\mathcal{E}(0)=-\frac{M}{2\pi}\int\limits_{-B}^{B}\cosh(\theta)\epsilon(\theta)d\theta,
\end{equation}
where the non-negative function $\epsilon(\theta)$ satisfies inside the interval $-B<\theta<B$, the BA equation
\begin{equation}\label{BA}
 \int\limits_{-B}^{B}\widetilde{K}(\theta-\theta^{\prime})\epsilon(\theta^{\prime})d\theta^{\prime}=A-M\cosh\theta
\end{equation}
and the parameter $B$ is determined by the boundary condition $\epsilon(\pm B)=0$. The kernel $\widetilde{K}(\theta)$ in this equation is related to the $\psi _{1}\psi_{1}$ scattering phase $\delta_{n,p}(\theta)$ \eqref{F} as
\begin{equation}\label{k1}
  \widetilde{K}(\theta)=\delta(\theta)-\frac{1}{2\pi i}\frac{d}{d\theta}\log(F(\theta))=\delta(\theta)-\frac{1}{2\pi}\frac{d}{d\theta}\delta_{n,p}(\theta).
\end{equation}
The Fourier transform $K(\omega)$ of  this kernel has the form
\begin{equation}\label{k2}
  K(\omega)=\frac{\sinh\left(\frac{\pi\omega}{n}\right)\cosh\left(\frac{\pi(p+n/2-1)\omega}{n}\right)}{\cosh(\frac{\pi\omega}{2})\sinh\left(\frac{\pi p\omega}{n}\right)}
\end{equation}
The main term of the asymptotics of the function $\mathcal{E}(A\rightarrow\infty)$ can be expressed explicitly through the kernel $K(\omega)$ by the relation \cite{Fateev:1992tk}
\begin{equation}\label{ass2}
  \mathcal{E}(A\rightarrow\infty)=-\frac{A^{2}}{2\pi K(0)}=-\frac{pA^{2}}{2\pi}.
\end{equation}
Comparing  \eqref{ass1} and \eqref{ass2} we find that
\begin{equation}\label{pb}
  p=\frac{1}{\lambda}=(1+b^{2})=a^{2},
\end{equation}
in agreement with perturbation theory. It means that for $(1+b^{2})>\frac{n}{2}$ the particles $M_{a}$ disappear from the spectrum. We see that at $b\rightarrow\infty$, $\lambda\rightarrow0$ and the $S-$matrix \eqref{Sb}, \eqref{fst2} tends to the scattering matrix of $O(n+2)-$ sigma model \cite{Zamolodchikov:1978xm}. So it is natural to call our QFT as the dual theory to deformed $O(n+2)-$ sigma model.

The term $\mathcal{E}(0)$ in \eqref{en} is the bulk vacuum energy of QFT \eqref{Fac}, \eqref{afd}. It can be expressed through the kernel $K(\omega)$ by the relation
\begin{equation}\label{e0}
  \mathcal{E}(0)=-\frac{M^{2}}{8}\left(K(\omega)\cosh\left(\frac{\pi\omega}{2}\right)|_{\omega=i}\right)^{-1}=\frac{M^{2}\sin\left(\frac{\pi p}{n}\right)}
  {8\sin\left(\frac{\pi}{n}\right)\sin\left(\frac{\pi(p-1)}{n}\right)}
\end{equation}
For small $b$ the bulk vacuum energy will be
\begin{equation}\label{e0b}
  \mathcal{E}(0)=\frac{M^{2}n}{8\pi b^{2}}+\frac{M^{2}}{8}\cot\left(\frac{\pi}{n}\right)  +O(b^{2}).
\end{equation}
The first term here is the contribution of Toda potential term in \eqref{Fac}, \eqref{afd} and gives $\frac{nM_{0}^{2}}{8\pi b^{2}}$. The second term comes from the renormalization of $M=M_{0}(b^{2})$ and from the contribution of the vacuum energies of $\left[\frac{n+2}{2}\right]$ free Dirac fermions and $\left[\frac{n+1}{2}\right]$ bosonic Toda particles with masses $M_{a}$ described above. The contribution from fermions and bosonic Toda particles  can be easily calculated (we note that (for odd $n$) Toda particle has mass $M$ and cancels the $1/2$ of the contribution of one fermion). It is
\begin{equation}\label{su}
   \delta\mathcal{E}(0) =-\frac{M^{2}}{8\pi}\sum\limits_{a=1}^{[n/2]}4\sin^{2}\left(\pi\frac{a}{n}\right)\log\left(  \frac{M_{a}^{2}}{M^{2}}\right)=
   -\frac{M^{2}}{8\pi}\sum\limits_{a=1}^{[n/2]}4\sin^{2}\left(\pi\frac{a}{n}\right)\log\left(4\sin^{2}\left(\pi\frac{a}{n}\right)\right)
\end{equation}
The exact relation between the physical mass $M$ and the parameter $M_{0}$ in the Lagrangian \eqref{Fac} can be calculated by the BA methods  \cite{Fateev:1993av,Zamolodchikov:1995xk} and has the form
\begin{equation}\label{mm0}
  M=M_{0}\frac{\Gamma\left(\frac{p}{n}\right)  2^{-(1+(-1)^{n})\frac{b^{2}}{n}}}{\Gamma\left(\frac{1}{n}\right)\Gamma\left(\frac{p+n-1}{n}\right)}=
  M_{0}\left(1+\left(\frac{\psi\left(\frac{1}{n}\right)-\psi(1)-(1+(-1)^{n})\log4}{n}\right)b^{2}+\dots\right).
\end{equation}
It is easy to derive from \eqref{e0b}, \eqref{su}, \eqref{mm0} the agreement with the perturbation theory and the sum rule
\begin{equation}
 \frac{1}{2\pi}\sum\limits_{a=1}^{[n/2]}\sin^{2}\left(\frac{\pi a}{n}\right)\log\left(4\sin^{2}\left(\frac{\pi a}{n}\right)\right)=-\frac{1}{8\pi}\left(\pi\cot\left(\frac{\pi}{n}\right)+
 2\left(\psi\left(\frac{1}{n}\right)-\psi(1)-(1+(-1)^{n})\log4\right)\right),
\end{equation}
where $\psi(z)=\Gamma^{\prime}(z)/\Gamma(z)$. This simple calculation confirms the general fact that after introduction of the Toda potential counterterm our QFTs do not have the UV divergences.

For $b\rightarrow\infty$ we rewrite
\[
  \mathcal{E}(0)=\frac{M^{2}\sin\left(\frac{\pi p}{n}\right)}{8\sin\left(  \frac{\pi}{n}\right)  \sin\left(  \frac{\pi(p-1)}{n}\right)}=\frac{M^{2}}{8}(\cot(\pi/n)+\cot(\pi b^{2}/n)).
\]
The second term oscillates around zero with period $n/b^{2}.$ It is reasonable to define the bulk ground state energy as the mean value over some interval much bigger then this period. With this definition we derive that bulk ground state energy of $O(n+2)$-sigma model: $\mathcal{E(}0\mathcal{)=}\frac{M^{2}}{8}\cot(\pi/n),$ what agrees with large $n$ expansion.

The integral BA equation can be studied using the generalized Winner-Hopf method \cite{Gan63}. We can go further in the UV ($\frac{A}{M}\gg1$) analysis of the function $\mathcal{E}(A)$. In particular the UV corrections to \eqref{ass2} have the form $\left(\frac{M}{A}\right)^{2\nu_{s}j}$, where the exponents $\nu_{s}$ are defined (with the factor $1/i$) by the
nearest zero to the real axis (for each series of zeroes) of the kernel \eqref{k2} in the upper half-plane. The zeroes at $\omega_{k}=ink,(\nu_{1}=n)$ correspond to the exponents independent on $b^{2}$. We call these exponents ``non-perturbative''. The zeroes at $\omega_{m}=i\frac{n(2m+1)}{2(p+n/2-1)},(\nu_{0}=\frac{n}{2(p+n/2-1)})$ we call the perturbative ones. For small $b$ the corresponding exponents $2\nu_{0}j$ $=\frac{jn}{(p+n/2-1)}$ appear after summation of logarithms in the perturbative series. As a result the UV expansion has the form
\begin{equation}\label{ser}
  \mathcal{E}(A)-\mathcal{E}(0)=-\mathcal{E}(0)-\frac{pA^{2}}{2\pi}\sum_{k=0}^{\infty }\left(\frac{M}{A}\right)^{2nk}f^{(k)}\left(\frac{M}{A}\right),
\end{equation}
where the functions $f^{(k)}(\frac{M}{A})$ are regular series in $\left(\frac{M}{A}\right)^{2\nu_{0}}$
\begin{equation}\label{serj}
  f^{(k)}\left(\frac{M}{A}\right)=\sum_{j=0}^{\infty}f_{j}^{(k)}\left( \frac{M}{A}\right)  ^{\frac{jn}{(p+n/2-1)}}. 
\end{equation}
The generalized WH method gives an iterative procedure for calculation of the coefficients $f_{j}^{(k)}$ in the UV expansion. All these coefficients can be expressed through the residues $a_{m}$ and $b_{k}$ of the function $\rho(\omega)=N(\omega)/N(-\omega)$, where $N(\omega)$ is the function analytical in the lower half plane which factorizes the kernel
$K(\omega)=(N(\omega)N(-\omega))^{-1}$. Namely $a_{m}=\textrm{res}\rho\left(\omega\right)|_{\omega=\omega_{m}}$ and $b_{k}=\textrm{res}\rho\left(\omega\right)|_{\omega_{k}}$.

Before considering the UV-asymptotics we say about non-perturbative terms which disappear in UV. The first such term $\delta_{1}\mathcal{E}(A)$ is $-\frac{pA^{2}}{2\pi}f_{0}^{(1)}\left(
\frac{M}{A}\right)^{2n}$. To understand the origin of this term we consider our action \eqref{ann} or \eqref{acd} with $\mathbf{U}_{odd}^{(m)}(\varphi,\phi)$ or $\mathbf{U}_{ev}^{(m)}(\varphi,\phi)$. The external field $A$  couples to the current $j_{0}=\psi_{1}^{\ast}\psi_{1}=\partial_{1}\phi_{1}$. One can find the first ``non-perturbative'' correction to the function $\mathcal{E}(A)$ expanding path integral near the massless point $\mu=\frac{M_{0}}{4\pi}$. The first correction is of order $A^{2}\left(\frac{M_{0}}{A}\right)^{2n}$. It corresponds to the minimal configuration of the exponents of the free massless fields $(\phi,\varphi)$ which gives a non-zero expectation value. The calculation of these contributions to $\mathcal{E}(A)$ reduces to the evaluation of the Coulomb-like integrals which can be calculated explicitly. Comparing this contribution with $\delta_{1}\mathcal{E}(A)$ derived from the BA equation we derive the exact relation between the parameters $M_{0}$ and $M$ \eqref{mm0}.

For $\frac{A}{m}\gg1$, the non-perturbative terms do not contribute and
\begin{equation*}
  \mathcal{E}(A)=-\frac{pA^{2}}{2\pi}\left(f^{(0)}\left(\frac{M}{A}\right)+O\left(\left(\frac{M}{A}\right)^{2n}\right)\right),
\end{equation*}
where $f^{(0)}\left(\frac{M}{A}\right)$ has  regular expansion in $\left(\frac{M}{A}\right)^{2\nu_{0}}$, $2\nu_{0}=\frac{n}{(p+n/2-1)}$
\begin{equation*}
   f^{(0)}\left(\frac{M}{A}\right)=\left(1+f_{1}^{(0)}\left(\frac{M}{A}\right)^{2\nu_{0}}+f_{2}^{(0)}\left(\frac{M}{A}\right)  ^{4\nu_{0}}+\dots\right).
\end{equation*}
For example the first coefficient is
\begin{equation*}
  f_{1}^{(0)}=-\frac{\left(2p-n+1\right)^{2}\Gamma\left(\frac{-1}{2p^{\prime}}\right)\Gamma\left(\frac{p}{2p^{\prime}}\right)\Gamma\left(\frac{1}{2}+
  \frac{n}{4p^{\prime}}\right)}{\left(p-n+\frac{1}{2}\right)^{2}\Gamma\left(\frac{1}{2p^{\prime}}\right)\Gamma\left(\frac{-p}{2p^{\prime}}\right) 
  \Gamma\left(  \frac{1}{2}-\frac{n}{4p^{\prime}}\right)}\left(\frac{\Gamma(\frac{n+1}{n})\Gamma(1+\frac{p-1}{n})}{\Gamma(1+\frac{p}{n})}\right)^{2\nu_{0}},
\end{equation*}
where $p^{\prime}=p+n/2-1$.  The next coefficients also can be found. The calculation simplifies drastically in the scaling limit $p\rightarrow\infty$, $\log\left(\frac{A}{M}\right)\rightarrow\infty$ with $\frac{\log\left(\frac{A}{M}\right)}{p}$ fixed. For example for $p\gg1$
\begin{equation*}
  f_{1}^{(0)}=-2\left(1+\frac{2+n-\log4+2n\log\Gamma(\frac{n+1}{n})}{2p}+O\left(\frac{1}{p^{2}}\right)\right).
\end{equation*}
In this limit the BA equations simplify and admit the exact solution. Corrections to the scaling behavior also can be systematically developed. Here
we quote the scaling limit of $\mathcal{E}(A)$ together with leading correction
\begin{equation}\label{scf}
  \mathcal{E}(A)=-\frac{pA^{2}}{2\pi}\frac{1-q}{1+q}\left(1+\frac{2}{p}\frac{q\log\left(\frac{(1-q^{2})p}{n}\right)}{1-q^{2}}+O\left(\frac{\log^{2}p}{p^{2}}\right)\right),
\end{equation}
where $q=\left(\frac{M}{A}\Gamma(\frac{n+1}{n})\left(\frac{e^{n+1/2}}{4}\right)^{1/n}\right)^{\frac{n}{p}}$. The main term 
\begin{equation}\label{scf-main}
  \mathcal{E}(A)=-\frac{pA^{2}}{2\pi}\frac{1-q}{1+q}+\dots=-\frac{pA^{2}}{2\pi}\tanh\left(\frac{n}{2p}\log\left(\frac{M}{A}\right)\right)+\dots,
\end{equation}
corresponds to one loop sigma-model metric. The next term should appear in the two-loop approximation. In the limit $p\rightarrow\infty$ we recover the result of \cite{Hasenfratz:1990ab} for $O\left(n+2\right)$ sigma model.

The function $\mathcal{E}(A)$ can be also calculated for $n\gg1$, $p\gg1$ with $\frac{p}{n}$ fixed, and arbitrary $\frac{A}{M}>1$ \cite{Fateev:1993av}. In this case
it can be written in the parametric form
\begin{equation}\label{w-s}
  \mathcal{E}(A)-\mathcal{E}(0)=-\frac{pA^{2}}{2\pi}\left(1-2X-\frac{n}{p}\left(X(1-X)(1-X^{\frac{2p}{n}}(1-X)^{-\frac{2p}{n}}\right)\right),\;
  \frac{M^{2}}{A^{2}}=4X^{\frac{2p+n}{n}}(1-X)^{-\frac{2p+n}{n}}.
\end{equation}
For $\frac{n}{p}\gg1,$ weak coupling, we derive from these relations
\begin{equation*}
  \mathcal{E}(A)-\mathcal{E}(0)=-\frac{pA^{2}}{2\pi}\left(\left(1-\frac{M^{2}}{A^{2}}\right)^{1/2}-\frac{M^{2}}{A^{2}}\log\left(\frac{A}{M}+
  \left(\frac{A^{2}}{M^{2}}-1\right)^{1/2}\right)\right)+O\Bigl(\frac{p}{n}\Bigr).
\end{equation*}
This function corresponds to almost free fermionic ground state energy. In the strong coupling limit $\frac{p}{n}\gg1$ we derive
\begin{equation*}
\mathcal{E}(A)-\mathcal{E}(0)=-\frac{pA^{2}}{2\pi}\left(\frac{1-\left(\frac{M}{A}\right)^{\frac{p}{n}}}{1+\left(\frac{M}{A}\right)^{\frac{p}{n}}}+O\left(\frac{n}{p}\right)\right),
\end{equation*}
that coincides with scaling approximation.
\section{Ricci flow}\label{Ricci-flow}
The non-linear sigma models (SM) in two space-time dimensions are widely used in QFT as well as in relation with string theory. They are described by the action
\begin{equation}\label{metr}
\mathcal{A}\left[  G\right]  =\frac{1}{4\pi}\int G_{ij}(X)\partial_{\mu}X^{i}\partial_{\mu}X^{j}d^{2}x+\dots
\end{equation}
where $X^{i}$ are coordinates in $d-$dimensional manifold called target space and symmetric matrix $G_{ij}(X)$ is the corresponding metric.  The terms denoted by $\dots$ can include topological terms, dilaton, $B$ field and higher forms consistent with the metric. 

The standard approach to two dimensional SM is the perturbation theory. If the curvature is small one can use the following renormalization group (RG)-evolution equation \cite{Friedan:1980jm}. Let $t$ be the RG time (the logarithm of the scale) which tends to $t\rightarrow-\infty$ in the UV limit and $t\rightarrow\infty$ in the IR. Then the one loop RG evolution equation reads
\begin{equation}\label{Ricci}
\frac{d}{dt}G_{ij}=-R_{ij}+O(R^{2})
\end{equation}
where $R_{ij}$ is the Ricci tensor for the metric $G_{ij}$.  In the case when the theory is not $\mathbf{P}$ and $\mathbf{T}$ invariant but only $\mathbf{PT}$ invariant this equation contains additional terms, related with $B-$ field.

The analysis of this equation shows \cite{Perelman:fk} that in general the non-linear evolution equation is unstable in the sense that even if one starts from manifold of  small curvature everywhere at some scale $t_{0}$, under evolution in both directions $t\rightarrow\pm\infty$ the metric $G_{ij}(t)$ develops at least some regions where its curvature grows and
\eqref{Ricci} is no more applicable. If it happens in the UV direction $t\rightarrow-\infty$ the action \eqref{metr} does not define  local QFT. However, special solutions exist where UV direction is stable and curvature remains small up $t\rightarrow-\infty$, permitting one to define the local QFT (at least perturbatively). For example, if we have homogeneous symmetric
space, its metric grows in the UV and curvature monotonously decreases and we are dealing with an UV asymptotically free QFT unambiguously defined by the action \eqref{metr}. Very interesting class of  solutions of Ricci flow equation form the solutions related with deformed symmetric spaces. The simple examples of such solutions are considered later. The asymptotic of  solutions of Ricci flow equations at $t\rightarrow-\infty$ correspond to the fixed points of  Ricci flow. They are more symmetric and subject to methods of CFT.

To study the large distance physics one should find a suitable approach. The quantum integrability is one of the most successful lines in studying of non-critical SMs. The quantum integrability and global symmetries of metric are manifested in the factorized scattering theory (FST) of corresponding excitations. The FST is rather rigid and its internal restriction does not permit a wide variety of consistent constructions. The FST contains all the information about background integrable QFT. The methods of integrable QFTs allows one to compute some off-mass-shell observables on the base of FST. In the UV region these observables should be compared with that's follows from SM \eqref{metr}. If they match non-trivially in the UV region it  naturally suggests the chosen FST as the scattering theory of integrable SM. Moreover, one can use the FST as a non-perturbative definition of the SM.

In this Section we consider the exact solutions of Ricci flow equations, namely the cases $n=1$ (sausage model \cite{Fateev:1992tk,Fateev:2017mug}) and $n=2$, corresponding to
the deformed spheres $S^{2}$  and $S^{3}$ and provide a generalization for $n>2$.

\paragraph{The case $n=1$.}
It is convenient to use coordinates different from the ones used in \cite{Fateev:1992tk,Fateev:2017mug}. These coordinate are more convenient for study the cases $S^{2}$ and $S^{3}$ at the same foot. In the first case the target space is two dimensional and we introduce coordinates $X^{1}=\vartheta$, where $-\pi/2\leq\vartheta<\pi/2$ and $X^{2}=\chi$, $0\leq\chi<2\pi$. The metric which solves the Ricci flow equation \eqref{Ricci} in these coordinate has the form
\begin{equation}\label{saus}
ds^{2}=\frac{\tanh(u)}{\nu}\left[\frac{1}{1-\tanh^{2}u\cos^{2}\vartheta}\bigl((d\vartheta)^{2}+\sin^{2}\vartheta(d\chi)^{2}\bigr)\right]
\end{equation}
Here and later $u=\nu(t_{0}-t)$ and $\nu$ plays the  role of deformation parameter. For small $\nu$ we can use  the perturbation theory in renormalization group (RG) equations. Here we use the main order approximation which is valid for small $\nu$. At $\nu\rightarrow0$ our metric tends to the metric of $S^{2}$. For $u\gg1$ the surface with the metric 
\eqref{saus} embedded into three dimensional space looks like a long sausage \cite{Fateev:1992tk}.

\paragraph{The case $n=2$.}
In this case we introduce the coordinates $X^{1}=\vartheta$, where $0\leq\vartheta<\pi$ and $X^{2}=\chi_{1}$, $X^{3}=\chi_{2}$, $0\leq\chi_{1,2}<2\pi$. The metric corresponding to the solution to \eqref{Ricci} in these coordinate has the form \cite{Fateev:1995ht}
\begin{equation}\label{3ds}
ds=\frac{\tanh\left(2u\right)  }{\nu}\left(  \frac{(d\vartheta)^{2}}{(1-k_{1}^{2}\sin^{2}\vartheta)(1-k_{1}^{2}\cos^{2}\vartheta)}+
\frac{\cos^{2}\vartheta(d\chi_{1})^{2}}{1-k_{1}^{2}\sin^{2}\vartheta}+\frac{\sin^{2}\vartheta(d\chi_{2})^{2}}{1-k_{1}^{2}\cos^{2}\vartheta}\right)
\end{equation}
where $k_{1}^{2}=\frac{2\sinh^{2}u}{\cosh2u}$. At $\nu\rightarrow0$ this metric tends to the metric of $S^{3}$. At $u\gg1$ the manifold with the metric \eqref{3ds} looks as the product of the sausage and the circle of radius $\sqrt{1/\nu}$.

\paragraph{The case $n>2$.}
Starting from $n=3$ the theory lacks $\mathbf{P}$ and $\mathbf{T}$ symmetry, but has only $\mathbf{PT}$ symmetry. It means that the non-zero $B$-field should obligatory appear in the action.  As an example, consider the case of $n=3$, which corresponds to the deformed sphere $S^{4}$. We introduce the coordinates $X^{1}=\vartheta_{1}$, $X^{2}=\vartheta_{2}$, where $-\pi/2\leq\vartheta_{1,2}<\pi/2$ and $X^{3}=\chi_{1}$, $X^{4}=\chi_{2}$, $0\leq\chi_{1,2}<2\pi$. Then the metric and the $B$-field are given by
\begin{equation}\label{metric-s4}
   ds^{2}=\frac{\kappa}{\nu}\Biggl[\frac{1}{(1-\kappa^{2}\cos^{2}\vartheta_{1})}\bigl(d\vartheta_{1}^{2}+\sin^{2}\vartheta_{1}(d\chi_{1})^{2}\bigr)+
   \frac{\cos^{2}\vartheta_{1}}{1-\kappa^{2}\cos^{4}\vartheta_{1}\cos^{2}\vartheta_{2}}\bigl((d\vartheta_{2})^{2}+\sin^{2}\vartheta_{2}(d\chi_{2})^{2}\bigr)\Biggr],
\end{equation}
and
\begin{equation}\label{Bfield-s4}
   B=-\frac{i\kappa^{2}\sin\vartheta_{2}\cos\vartheta_{2}\cos^{4}\vartheta_{1}}{\nu(1-\kappa^{2}\cos^{4}\vartheta_{1}\cos^{2}\vartheta_{2})}d\vartheta_{2}\wedge d\chi_{2}.
\end{equation}
The metric \eqref{metric-s4} and the $B-$field \eqref{Bfield-s4} satisfy modified Ricci flow equations
\begin{equation}\label{modified-Ricci}
      R_{ij}-\frac{1}{4}H_{i}^{kl}H_{jkl}+\nabla_{i}V_{j}+\nabla_{j}V_{i}=-\dot{G}_{ij},\quad
      H_{k ij}V^{k}-\frac{1}{2}\nabla_{k}H^{k}_{ij}+\nabla_{i}\omega_{j}-\nabla_{j}\omega_{i}=-\dot{B}_{ij},
\end{equation}
provided that  the vector $V_{i}$ and the one-form $\omega$ are given by
\begin{equation*}
  V_{i}=\left(\frac{2\kappa^2 \sin\vartheta_{1}\cos\vartheta_{1}(1-\cos^{2}\vartheta_{1}\cos^2\vartheta_{2})}
  {(1-\kappa^2\cos^2\vartheta_{1})(1-\kappa^2\cos^4\vartheta_{1}\cos^2\vartheta_{2})},0,0,0\right),\quad
  \omega=\frac{i\kappa\cos^{2}\vartheta_{1}\sin^{2}\vartheta_{2}}{(1-\kappa^{2}\cos^{4}\vartheta_{1}\cos^{2}\vartheta_{2})}d\chi_{2},
\end{equation*}
and the coupling $\kappa=\kappa(t)$ satisfies differential equation
\begin{equation}\label{kappa-diff}
   \dot{\kappa}=3\nu(\kappa^{2}-1).
\end{equation}
Differential equation \eqref{kappa-diff} has a solution $\kappa=\tanh3u$. We note that that acting with $t$-dependent diffeomorphism and making the gauge transformation, one can always reduce modified Ricci equation \eqref{modified-Ricci}  to the more standard form where the vector $V_{i}$ and the one-form $\omega$ vanish. In the limit $\nu\rightarrow0$ the metric \eqref{metric-s4} approaches the metric of the round four-sphere, while the $B-$field, the vector $V_{i}$ and the one-form $\omega$ tend to zero. 

It is interesting to apply $T-$duality transformation to the $\chi_{2}$ isometry direction. The $B-$field will vanish, while the metric become complex and will take the form
\begin{multline}\label{metric-s4-dual} 
   ds^{2}=\frac{\kappa}{\nu}\Biggl[\frac{d\vartheta_{1}^{2}+\sin^{2}\vartheta_{1}(d\chi_{1})^{2}}{(1-\kappa^{2}\cos^{2}\vartheta_{1})}+\cos^{2}\vartheta_{1}(d\vartheta_{2})^{2}+\\+
   2i\cos^{2}\vartheta_{1}\cot\vartheta_{2}d\vartheta_{2}d\chi_{2}+
   \frac{1-\kappa^{2}\cos^{4}\vartheta_{1}\cos^{2}\vartheta_{2}}{\kappa^{2}\cos^{2}\vartheta_{1}\sin^{2}\vartheta_{2}}(d\chi_{2})^{2}\Biggr],
\end{multline}
This metric satisfies
\begin{equation}\label{RG-equation-dilaton}
  R_{ij}+2\nabla_{i}\nabla_{j}\Psi=-\dot{G}_{ij},\qquad
  |\nabla \Psi|^{2}-\frac{1}{2}\Delta\Psi=-\dot{\Psi},
\end{equation}
with the dilaton field
\begin{equation}\label{S4-dilaton}
   \Psi=\frac{1}{2}\log\left(\frac{(1-\kappa^{2}\cos^{2}\vartheta_{1})^{2}}{\kappa(1-\kappa^{2})\cos^{2}\vartheta_{1}\sin^{2}\vartheta_{2}}\right)-i\chi_{2}.
\end{equation}
We note that the metric \eqref{metric-s4}-\eqref{Bfield-s4} and the dual metric \eqref{metric-s4-dual} appear also in \cite{Arutyunov:2013ega,Hoare:2015gda}. The metric in \cite{Arutyunov:2013ega,Hoare:2015gda} has been derived from the general construction suggested in \cite{Klimcik:2008eq,Delduc:2013fga}. 
\section{Minisuperspace considerations}\label{MSS}
Here we perform some calculations to support the identification of QFT's described by the SM's \eqref{saus}, \eqref{3ds}  and \eqref{metric-s4}-\eqref{Bfield-s4} with the FST's \eqref{Sb}, \eqref{fst2}. First we study the infinite volume specific vacuum energy $\mathcal{E}(A)$ in the constant external field $A$. The action \eqref{metr} is invariant with respect to global axial rotation $X^{2}\rightarrow X^{2}+\zeta$ for $n=1$ and $n=2$. It means that we can couple our QFT's with constant field $A$ by transformation $\partial_{0}\rightarrow\partial_{0}-iA.$ This field introduces the scale and we can assign
\begin{equation*}
   (t_{0}-t)=\log\left(\frac{A}{M}\right),\quad u=\nu\log\left(\frac{A}{M}\right).
\end{equation*}
To ensure the validity of our one loop action we go to the scaling limit $\nu\rightarrow0$, $\frac{A}{M}\rightarrow\infty$ with the scaling variable $u=\nu\log\left(\frac{A}{M}\right)$ fixed. The ground state energy $\mathcal{E}(A)$ is equal to the minimum of the density of the action \eqref{metr} with $\partial_{0}\rightarrow\partial_{0}-iA$. This minimum is
achieved at $\vartheta=0$ and is equal to
\begin{equation}\label{sce}
  \mathcal{E}_{n}(A)=-\frac{A^{2}}{4\pi\nu}\tanh(nu),\quad\text{for}\quad n=1,2.
\end{equation}
The same result is valid for all $n$. The next (two-loop) correction is of relative order $\nu\log\nu$.

Comparing this result with \eqref{scf} and \eqref{scf-main} derived from BA equations we find that in the scaling limit they coincide and 
\begin{equation*}
   \nu=\frac{1}{2p}=\frac{1}{2\left(1+b^{2}\right)}=\frac{1}{2a^{2}}
\end{equation*}
and we see that loop expansion is the strong coupling expansion for our QFT's.

The second observable that we consider is the energy levels $E_{i}(R)$ of our SMs at the circle of length $R.$ It introduces the scale $(t_{0}-t)=\log\left(\frac{1}{RM}\right)$. In the UV scaling regime: $-\log(RM)\rightarrow\infty$, $\nu\rightarrow0$ such that $u=-\nu\log(RM)$ is finite, our one-loop approximation is exact up to $O(\nu\log\nu)$. In this approximation we can use minisuperspace approach to calculate the UV corrections to Ground State Energy $E_{0}(R)$, effective central charge $E_{0}(R)=-\frac{\pi}{6R}c(R)$ and to the energies $E_{i}(R)$ of the exited states. It was in shown in \cite{Fateev:1992tk} that these values can be expressed throw the eigenvalues of the covariant operator $\hat{h}$
\begin{equation}\label{mss}
  \hat{h}=-\nabla_{t}^{2}+\frac{1}{4}\mathcal{R}_{t},\quad\hat{h}\Psi_{i}=\frac{e_{i}(R)}{6}\Psi_{i}
\end{equation}
where $\nabla_{t}^{2}$ is the Laplace operator and $\mathcal{R}_{t}$ is the scalar curvature in the SM metric renormalized at the scale $R$. Then with the accuracy $\nu\log\nu$
\begin{equation}\label{m1}
   c(R)=2-e_{0}(R),\quad E_{i}(R)=E_{0}(R)+\frac{\pi(e_{i}(R)-e_{0}(R))}{R}.
\end{equation}
Operator $\hat{h}$  is self-adjoint with respect of scalar product with the SM metric
\begin{equation*}
  \left(\Psi_{1},\Psi_{2}\right)=\int\Psi_{1}^{\ast}\Psi_{2}\sqrt{\det G}\,dx^{1}\dots dx^{n+2},
\end{equation*}
where coordinates $x^{i}$ can be considered as the zero modes of fields $X^{i}$. It is easy to see that operator $\hat{h}$  depends only on the scaling variable $u=-\nu\log(RM)$. It means that the eigenvalues $e_{i}(R)$ scale as $\nu e_{i}(u)$. Here we consider the cases $n=1,2$ separately.
\paragraph{The case $n=1$.}We can search for the solution $\Psi=e^{ix_{2}m}\Psi_{m}$. After the substitution
\[
 \frac{\sin^{2}\vartheta}{1-\tanh^{2}u\cos^{2}\vartheta}=sn^{2}(z|s),\quad \psi_{m}=\sqrt{sn(z|s)}\Psi_{m},
\]
with the modulus of elliptic Jacobi function $s^{2}=\tanh^{2}u$  the minisuperspace equation can be written in the Lam\'{e} form
\begin{equation}\label{lame}
  \left(-\frac{d^{2}}{dz^{2}}-\frac{cn^{2}(z|s)dn^{2}(z|s)}{4sn^{2}(z|s)}+\frac{m^{2}}{sn^{2}(z|s)}\right)\psi_{m}=\frac{s\,\kappa_{m,j}}{6}\psi_{m},
\end{equation}
where $e_{m,j}(R)=\nu\kappa_{m,j}(u)$ and the boundary conditions for the solutions are: $\psi_{m}\sim z^{m+\frac{1}{2}}$ at $z\rightarrow0,\psi_{m}\sim(K-z)^{m+\frac{1}{2}}$ at $z\rightarrow K$, where $K\left(s^{2}\right)$ is the real period of Jacobi functions. For arbitrary $u$ or $s^{2}=\tanh^{2}u$ this equation can be solved numerically but for small and large $u$ it admits analytical solution.

For small $u$, $s\simeq u$, $K\simeq\frac{\pi}{2}$, the equation \eqref{lame} can be easily solved: $\psi_{m}=(\sin z)^{1/2}P_{j}^{(m)}(\cos z)$, where $P_{j}^{(m)}$ are the Legendre spherical functions. We derive $\frac{\kappa_{m,j}}{6}\simeq\frac{j\left(  j+1\right)+1/2}{u}$; $j\geq m$ and $e_{0}(R)=\nu\kappa_{m,0}=\frac{3}{\log(1/RM)}$. This asymptotic is
universal for all spheres $S^{d}$ with $d>1$
\begin{equation}\label{cd}
 c(R)=d-\frac{3}{2}d/\log(1/RM)+O\left(\log\left(\log\left(1/RM\right)\right)/\log^{2}(1/RM)\right).
\end{equation}
It is interesting to calculate the first corrections to the levels at small $u$
\begin{equation}\label{k00}
\begin{aligned}
  &\frac{\kappa_{m,j}}{6}=\frac{j\left(j+1\right)+1/2}{u}-\frac{2\left(j\left(j+1\right)-1\right)(j\left(j+1\right)-3m^{2})}{3\left(2j-1\right)\left(  2j+3\right)}u+O(u^{3}),\\
  &\frac{\kappa_{0,0}}{6}=\frac{1}{2}u-\frac{2}{135}u^{3}+O(u^{5})
\end{aligned}
\end{equation}
We see that three first levels (corresponding to three particles in IR) with the quantum numbers $j=1$, $m=\pm1$, split moving to UV.

We consider now another limit $u\gg1$, $s^{2}\rightarrow1$, $K\simeq u+\log2$. In this limit the potential $V(z)$ in the Lame equation with exponential accuracy looks as
\begin{equation}\label{V}
\begin{aligned}
  V_{l}(z)&=-\frac{1}{\sinh^{2}2z}+m^{2}\coth^{2}z\quad0<z\ll K,\\
  V_{r}\left(z_{1}\right)   &  =-\frac{1}{\sinh^{2}2z_{1}}+m^{2}\coth^{2}z_{1}\quad0<z_{1}=K-z\ll K.
\end{aligned}
\end{equation}
We parametrize $\frac{\kappa_{m,n}}{6}=m^{2}+4p^{2}$. Then in the middle one can neglect the potential term and $\psi_{m}$ is the plane wave solution. At the left and right ends $z\sim0,z\sim K$  the equation can be solved exactly in terms of hypergeometric functions $F(A,B,C,z)$
\begin{align}\label{ps}
  \psi_{m}^{(l)}  &  =N_{l}\left(p,m\right)\left(\tanh z\right)^{m+\frac{1}{2}}\left(\cosh z\right)^{2ip}F\left(A,A,m+1,\tanh^{2}z\right)\nonumber\\
  \psi_{m}^{(r)} &  =\psi_{m}^{(l)}(z_{1}),\quad\text{where}\quad A=\frac{m+1-2ip}{2}
\end{align}
The constant $N_{l}(p,m)=N_{r}(p,m)$ is chosen from the condition
\begin{equation*}
  \psi_{m}^{(l)}\simeq e^{2ipz}+\mathrm{R}_{l}^{(\textrm{cl})}(p,m)e^{-2ipz},\quad
  \psi_{m}^{(r)}\simeq e^{i2pz_{1}}+\mathrm{R}_{r}^{(\textrm{cl})}(p,m)e^{-2ipz_{1}},\quad z\gg1.
\end{equation*}
The corresponding solutions \eqref{ps} are specified by the reflection amplitudes
\begin{equation}\label{rlr}
  \mathrm{R}_{l}^{(\textrm{cl})}=\mathrm{R}_{r}^{(\textrm{cl})}=\frac{\Gamma(1+2ip)\Gamma^{2}(\frac{1+|m|}{2}-ip)}{\Gamma(1-2ip)\Gamma^{2}(\frac{1+|m|}{2}+ip)}.
\end{equation}
Matching the solutions in different domains with a plane wave in the middle we derive
\begin{equation}\label{spec}
  \frac{1}{6}\kappa_{m,n}=m^{2}+\frac{\pi^{2}(j+1)^{2}}{4(u+r_{m})^{2}}+O\left(\frac{1}{u^{5}}\right);\quad r_{m}=\psi(1)-\psi\left(\frac{m+1}{2}\right).
\end{equation}
In particular, the effective central charge $c(R)=2-e_{0}(R)$ is
\begin{equation*}
   c(R)=2-e_{0}(R)=2-6\nu\kappa_{0,0}(u)=2-\frac{3\pi^{2}b^{2}}{\left(\log\left(\frac{1}{MR}\right)+b^{2}4\log2\right)^{2}}
\end{equation*}
The relative correction to $c(R)$ is of order $\frac{\log b^{2}}{b^{2}}$. The comparison of the scaling function $k_{0,0}(u)$ derived from the numerical solution of equation \eqref{lame} with the same scaling function derived from TBA equations \cite{Fateev:1992tk} gives an excellent agreement. 
\paragraph{The case $n=2$.} In this case we can search for the solution $\Psi=e^{ix_{2}m_{1}+ix_{3}m_{3}}\Psi_{m_{1}m_{2}}$. After elliptic substitutions the minisuperspace equation for the function $\psi_{m_{1}m_{2}}=sn(z|s^{\prime})^{\frac{1}{2}} \Psi_{m_{1}m_{2}}$ has a form
\begin{equation}\label{lam1}
  \left(-\frac{d^{2}}{dz^{2}}+U_{m_{1}m_{2}}(z|s^{\prime2})\right)\psi_{m_{1}m_{2}}=\frac{s^{\prime}\kappa_{m_{1},m_{2},j}^{\prime}}{6}\psi_{m_{1}m_{2}},
\end{equation}
where the potential term $U_{m_{1}m_{2}}(z|s^{\prime2})$ is
\begin{equation*}
  -\frac{cn^{2}(z|s^{\prime})dn^{2}(z|s^{\prime})}{4sn^{2}(z|s^{\prime})}-\frac{(1-s^{\prime2})sn^{2}(z|s^{\prime})}{4cn^{2}(z|s^{\prime})dn^{2}(z|s^{\prime})}+
  \frac{m_{1}^{2}}{sn^{2}(z|s^{\prime})}+\frac{m_{2}^{2}dn^{2}(z|s^{\prime})}{cn^{2}(z|s^{\prime})}%
\end{equation*}
with the modulus of elliptic Jacobi function $s^{\prime2}=\tanh^{2}2u$.  We note that the potential $U_{m_{1}m_{2}}(z|s^{2})$ is invariant under $z\rightarrow z+\frac{1}{2}K\left(s^{\prime2}\right)$, $m_{1}\rightarrow m_{2}$.

For small $u$, $s^{\prime}\simeq2u,$ $K\simeq\frac{\pi}{2}$, the equation \eqref{lam1} can be solved
\begin{equation*}
  \psi_{m_{1}m_{2}}=\sin^{m_{1}}(z)\cos^{m_{2}}(z)P_{j}^{\left(m_{1},m_{2}\right)}(\cos(2z))+O(u^{2})
\end{equation*}
and with $J=2j+|m_{1}|+|m_{2}|$
\begin{equation}
\begin{aligned}
  &\frac{\kappa_{m_{1},m_{2},j}^{\prime}}{6}=\frac{J(J+2)+3/2}{2u}-\frac{J(J+2)-6(m_{1}^{2}+m_{2}^{2})}{6}u+O(u^{3}),\\
  &\frac{\kappa_{0,0,0}^{\prime}}{6}=\frac{3}{4u}-\frac{8}{15}u^{3}+O(u^{5}).
\end{aligned}
\end{equation}
In this case due to the symmetry of the potential we have no the splitting of the levels with quantum numbers $j=1$, $m_{1}=\pm1$, $m_{2}=\pm1$.

We consider now the limit $u\gg1$, $s^{\prime2}\rightarrow1$, $\frac{K}{2}\simeq u+\frac{1}{2}\log2$. In this limit the the potential $U_{m_{1}m_{2}}(z|s^{\prime2})$ in the Lame equation with exponential accuracy looks as
\begin{equation}
\begin{aligned}
  &U_{l}(z)=-\frac{1}{\sinh^{2}2z}+m_{1}^{2}\coth^{2}z+m_{2}^{2},\quad0<z\ll\frac{K}{2},\\
  &U_{r}(z_{1})=-\frac{1}{\sinh^{2}2z_{1}}+m_{2}^{2}\coth^{2}z_{1}+m_{1}^{2},\quad0<z_{1}=\frac{K}{2}-z\ll\frac{K}{2}.
\end{aligned}
\end{equation}
We parametrize $\frac{\kappa_{m,n}}{6}=m_{1}^{2}+m_{2}^{2}+4p^{2}$. Then in the middle one can neglect the potential term and $\psi_{m_{1}m_{2}}$ is the plane wave solution. At the left and right ends $z\sim0$, $z_{1}\sim\frac{K}{2}$ the equation can be solved exactly in terms of hypergeometric functions
\begin{equation}
  \psi_{m_{1}m_{2}}^{(l)}=N_{l}(p,m_{1})(\tanh z)^{m_{1}+\frac{1}{2}}(\cosh z)^{2ip}F(A_{1},A_{1},m_{1}+1,\tanh^{2}z),
\end{equation}
where $A_{1}=\frac{m+1-2ip}{2}$ and the solution $\psi_{m_{1}m_{2}}^{(r)}=\psi_{m_{2}m1}^{(l)}(z_{1})$. These solutions are specified by the reflection amplitudes
\begin{equation}
   \mathrm{R}_{l}^{(\textrm{cl})}=\frac{\Gamma(1+2ip)\Gamma^{2}(\frac{1+|m_{1}|}{2}-ip)}{\Gamma(1-2ip)\Gamma^{2}(\frac{1+|m_{1}|}{2}+ip)},\,
   \mathrm{R}_{r}^{(\textrm{cl})}=\frac{\Gamma(1+2ip)\Gamma^{2}(\frac{1+|m_{2}|}{2}-ip)}{\Gamma(1-2ip)\Gamma^{2}(\frac{1+|m_{2}|}{2}+ip)}.
\end{equation}
Matching the solutions in different domains with a plane wave in the middle we derive
\begin{equation}
  \frac{\kappa_{m_{1},m_{2},j}^{\prime}}{6}=m_{1}^{2}+m_{2}^{2}+\frac{\pi^{2}(j+1)^{2}}{4(u+\frac{1}{2}(r_{m_{1}}+r_{m_{2}}+\log2))^{2}}.
\end{equation}
In particular, the effective central charge $c(R)=3-e_{0}(R)$ is
\begin{equation*}
  c(R)=3-6\nu\kappa_{0,0,0}(u)=3-\frac{3\pi^{2}b^{2}}{\left(\log\left(\frac{1}{MR}\right)+3b^{2}\log2\right)^{2}}
\end{equation*}
The relative correction to $c(R)$ is of order $\frac{\log b^{2}}{b^{2}}$. The numerical calculation of this function from TBA equations derived in \cite{Fateev:1996ea} has been done by E. Onofri.
\paragraph{The case $n=3$.}
In this case the metric depends on two variables and calculations becomes more involved. Even in the previous cases the minisuperspace equation for the Ricci flow of the levels could be solved only numerically. However the most interesting information which can be studied analytically,  comes from the UV limit $u\rightarrow\infty$ of these equations (for small $u\ll1$ the spectrum can be derived by usual perturbation theory in $u^{2}$). To derive the conformal limit of the metric and the $B-$field it is useful to use the method of contraction. At $u\rightarrow\infty$, $\kappa(u)=\tanh3u\rightarrow1$. We introduce the parameter $\delta/2=1-\tanh3u\rightarrow0$ and make the rescaling of variables $\vartheta_{1}\rightarrow\sqrt{\delta}\vartheta_{1}$, $\vartheta_{2}\rightarrow\sqrt{\delta}\vartheta_{2}$ tending $\delta\rightarrow0$. As a result the action can be written in terms of two complex scalar fields ($\nu_{k}=\vartheta_{k}e^{i\chi_{k}}$)
\begin{equation}\label{aa3}
   \mathcal{A}_{CFT}=\frac{1}{4\pi\nu}\int\left(\frac{\left(\partial_{\alpha}\nu_{1}\partial_{\alpha}\nu_{1}^{\ast}\right)}{1+\nu_{1}\nu_{1}^{\ast}}+
   \frac{(\delta_{a\beta}+\varepsilon_{\alpha\beta})\left(\partial_{\alpha}\nu_{2}\partial_{\beta}\nu_{2}^{\ast}\right)}{\left(1+2\nu_{1}\nu_{1}^{\ast}+
   \nu_{2}\nu_{2}^{\ast}\right)}\right)d^{2}x.
\end{equation}
The generalization of this action for arbitrary odd $n=2m-1$ will have the form
\begin{equation}\label{aan}
  \mathcal{A}_{CFT}=\frac{1}{4\pi\nu}\int\left(\frac{\left(\partial_{\alpha}\nu_{1}\partial_{\alpha}\nu_{1}^{\ast}\right)}{1+\nu_{1}\nu_{1}^{\ast}}+
  \sum_{j=2}^{m}\frac{(\delta_{a\beta}+\varepsilon_{\alpha\beta})\left(\partial_{\alpha}\nu_{j}\partial_{\beta}\nu_{j}^{\ast}\right)}
  {\left(1+2\nu_{1}\nu_{1}^{\ast}+\dots+2\nu_{j-1}\nu_{j-1}^{\ast}+\nu_{j}\nu_{j}^{\ast}\right)}\right)d^{2}x.
\end{equation}
For even $n=2m$ we should add one free field real field $\nu_{0}$ and the term $\frac{(\partial_{\alpha}\nu_{0})^{2}}{4\pi\nu}$ to the action. The actions \eqref{aa3}, \eqref{aan} possess besides the metric part also the $B-$field term, which modifies the minisuperspace equation \eqref{mss}. With this term this equation will have a form
\begin{equation}\label{HT}
  \left(-\nabla^{2}+\frac{1}{4}\mathcal{R}+\frac{1}{12}\mathcal{H}^{2}\right)\Psi_{i}=\frac{e_{i}}{6}\Psi_{i},
\end{equation}
where the $3$-form $\mathcal{H}$ is defined throw two form $B$ as  $\mathcal{H}=dB$.

In this paper we consider the relation between CFTs \eqref{ac2} and \eqref{aa3} for $k=2$. Namely, we conjecture that they are dual. For $k=1$ we have the known duality between Sine-Lioville and Witten black hole  CFTs. For $k=2$ we show that the reflection amplitudes of these CFTs coincide in the minisuperspace approximation. 

It is convenient to parametrize the continuous spectrum (in the conformal limit) spectrum of the \eqref{HT} as $\frac{e_{i}}{6}=P^{2}+Q^{2}$. The function $\Psi_{i}(\det G)^{1/4}$ which are integrated with trivial measure we denote as $\Psi_{P,Q}$.

The Weyl group $w_{2}$ of the Lie algebra of $C_{2}$ contains $8$ elements and acts to the parameters $P$ and $Q$ as independent transformations $P\rightarrow\pm P$, $Q\rightarrow\pm Q$, and $P\leftrightarrow Q$. We denote $P_{s}$, $Q_{s}$ the components of vector $(P,Q)$ after the action of the element $\hat{s}\in w_{2}$. For small $\nu_{1}$, $\nu_{2}$ the action \eqref{aa3} is close to the action of free fields. We parametrize $\nu_{1}=e^{x+i\chi_{1}}$, $\nu_{2}=e^{y+i\chi_{2}}$. The dependence on $\chi_{1}$ and $\chi_{2}$ is rather trivial, it leads to the multiplication of the wave function by the factor $e^{im_{1}\chi_{1}}e^{im_{2}\chi_{2}}.$  To simplify the notations and equations we consider here the ``isotopic symmetric'' sector where $\Psi_{P,Q}=\Psi_{P,Q}(x,y)$.

In the  ``Weyl chamber'' $x,y,y-x\ll0$, the function $\Psi_{P,Q}(x,y)$ can be represented as the Weyl superposition with numerical coefficients of functions $\Omega_{P_{s},Q_{s}}$, where the function $\Omega_{P},_{Q}$ is 
\begin{equation}
  \Omega_{P,Q}(x,y)=e^{iPx+iQy}R_{P,Q}(x,y)=e^{iPx+iQy}\left(1+\sum_{n_{1},n_{2},n_{3}>0}r_{n_{1},n_{2},n_{3}}(e^{x})^{n_{1}}(e^{-x+y})^{n_{2}}(e^{y})^{n_{3}}\right)
\end{equation}
All coefficients $r_{n_{1},n_{2},n_{3}}$ can be derived from the results of appendix \ref{mss-diff}. The coefficients $A(P_{s},Q_{s})$ before the functions $\Omega
_{P_{s},Q_{s}}(x,y)$ in the Weyl sum are directly related to the reflection amplitudes $\mathrm{R}_{\hat{s}}^{(\textrm{cl})}(P,Q)$, which are the special limit of the quantum reflection amplitudes \eqref{ra}, \eqref{ra1}. Namely in the limit $b\rightarrow\infty$, we set $b\mathrm{a}=b(\mathrm{a}_{1},\mathrm{a}_{2})=(iP,iQ)$. In isotopic symmetric sector we should put also $\mathrm{b}=0$.
\begin{equation}\label{racls}
  \mathrm{R}_{\hat{s}}^{(\textrm{cl})}(P,Q)=\frac{\mathrm{A}^{(\textrm{cl})}(P_{s},Q_{s})}{\mathrm{A}^{(\textrm{cl})}(P,Q)}
\end{equation}
where
\begin{equation*}
  \mathrm{A}^{(\textrm{cl})}(P,Q)=\frac{\Gamma(-2iQ)\Gamma(-2iP)\Gamma(-iQ-iP)\Gamma(iP-iQ)\Gamma(\frac{1}{2}+iP)\Gamma(\frac{1}{2}+iQ)^{2}}
  {\Gamma(\frac{1}{2}-iQ)^{2}\Gamma(\frac{1}{2}-iP)}.
\end{equation*}
The statement is that function $\Psi_{P,Q}(x,y)$, written in the symmetric form
\begin{equation*}
  \Psi_{P,Q}(x,y)=\sum_{w2}A(P_{s},Q_{s})\Omega_{P_{s}},_{Q_{s}}(x,y),
\end{equation*}
has correct properties if and only if  $A(P,Q)=\mathrm{A}^{(cl)}(P,Q)$. We show this in appendix \ref{mss-diff}.

We considered here the conformal limit of \eqref{HT}. However using this result we can derive the asymptotic of the energy levels for large values of $b^{2}$ and $\log(\frac{1}{mR})$ as it was done in \cite{Ahn:2000kia}. Using the method based on reflection amplitudes we derive that in the isotopic symmetric sector the asymptotic of the levels depends on two quantum numbers and has the form
\begin{equation*}
  e_{n_{1},n_{2}}(R)=\frac{3\pi^{2}b^{2}(5+(n_{1}+n_{2})((n_{1}+n_{2})+3)+n_{2}(n_{2}+1))}{4\left(\log\left(\frac{1}{MR}\right)+2b^{2}\log2\right)^{2}}.
\end{equation*}

At the end of this Section we discuss the QFT with the $T-$dual metric \eqref{metric-s4-dual}. It is the solution of Ricci flow equations with $\kappa(u)$ satisfying the
equation \eqref{kappa-diff}. We can take $\kappa(u)=\coth3u$. As it was shown in \cite{Fateev:2017mug} (for $n=1$, and is valid for any $n$), besides  integrable QFT with the action \eqref{ann} we can consider another integrable QFT with the action where for every pair $\{e^{ia\phi_{m}+b\varphi_{m-1}},e^{-ia\phi_{m}-b\varphi_{m}}\}$ we can introduce the ``dual'' field $\hat{\phi}_{m}:\partial_{\alpha}\hat{\phi}_{m}=\varepsilon_{\alpha,\beta}\partial_{\beta}\phi_{m}$. The corresponding action will be well defined only if the terms $e^{ia\phi_{m}+b\varphi_{m-1}}$ and $e^{-ia\hat{\phi}_{m}-b\varphi_{m}}$ in the pair $\{e^{ia\phi_{m}+b\varphi_{m-1}},e^{-ia\hat{\phi}_{m}-b\varphi_{m}}\}$ will be mutually local.
It happens if 
\begin{equation*}
   2a^{2}=\mathrm{p},\quad\text{where}\;\mathrm{p}\text{ is an integer}.
\end{equation*}
For $\mathrm{p}\gg1$ the one loop approximation to the SM representation with the $T-$dual metric works and corrections are of order $\frac{1}{p}\log p$.

In the UV limit $\kappa(u)=\coth3u=1+\frac{\delta}{2}$. We take the local coordinates $\vartheta_{1}\rightarrow i\sqrt{\delta}\vartheta_{1},\vartheta_{2}\rightarrow i\sqrt{\delta}\vartheta_{2}$ and derive 
\begin{equation}\label{mUV}
  ds_{UV}=\mathrm{p}\left(  \frac{(d\vartheta_{1})^{2}}{1+\vartheta_{1}^{2}}+\frac{\vartheta_{1}^{2}(d\chi_{1})^{2}}{1+\vartheta_{1}^{2}}
   +\frac{2id\vartheta_{2}d\chi_{2}}{\vartheta_{2}}+\frac{(1+2\vartheta_{1}^{2}+\vartheta_{2}^{2})(d\chi_{2})^{2}}{\vartheta_{2}^{2}}\right).
\end{equation}
We note that with this metric we derive exactly the same equation to the function $\Psi_{P,Q}$ (in this case $B=0$ and term $\mathcal{H}^{2}$ disappears) but only in the isotopic symmetric sector.

In the IR limit $\coth3u=-(1+\frac{\delta}{2})$ and we take the local coordinates $\vartheta_{1}\rightarrow\sqrt{\delta}\vartheta_{1},\vartheta_{2}\rightarrow\sqrt{\delta}\vartheta_{2}$ and derive 
\begin{equation}\label{mIR}
  ds_{IR}=\mathrm{p}\left(  \frac{(d\vartheta_{1})^{2}}{1-\vartheta_{1}^{2}}+\frac{\vartheta_{1}^{2}(d\chi_{1})^{2}}{1-\vartheta_{1}^{2}}
            +\frac{2id\vartheta_{2}d\chi_{2}}{\vartheta_{2}}+\frac{(1-2\vartheta_{1}^{2}-\vartheta_{2}^{2})(d\chi_{2})^{2}}{\vartheta_{2}^{2}}\right),
\end{equation}
and we see that now coordinates change in the finite domain and minisuperspace equation has the discrete spectrum. It means that we have the RG flow from UV to other critical critical point in the IR, which is described by rational CFT. This phenomenon was discussed in details in \cite{Fateev:2017mug} for the case $n=1$. Here we discuss this situation in the Concluding Remarks.
\section{Concluding remarks}\label{concl}
In this Section we discuss some important points which were missed in the main
body of the paper.
\begin{enumerate}
\item In the expansion for the specific ground state energy in the external field $A$ \eqref{ser} we had two types of contributions: the ``non-perturbative'' terms $\left(\frac{M}{A}\right)^{2nk}$ and ``perturbative'' ones $\left(\frac{M}{A}\right)^{\frac{nj}{(p+n/2-1)}}$. The origin of the ``non-perturbative'' terms can be easily explained from the actions \eqref{ann} and \eqref{acd} expressed in terms of  fields $\left(\varphi_{i},\phi_{i}\right)$. The perturbative can be easily explained in the dual SM representation for our QFT.  The appearance of ``non-perturbative'' terms in the SM representation was established in details in \cite{Fateev:1992tk} for the case $n=1$ (sausage model), where they were related with $k-$instantons contributions which had the topological origin. For arbitrary integer $n>1$ we have no topological arguments, but the action of the deformed $O(n+2)$ SM always contains the part corresponding to sausage model. It means that we have the particular solutions corresponding to embedding of $k-$instantons solutions. The SM action on these solutions gives ``non-perturbative'' contributions to the observables in the SM-representation.
\item The $W-$ algebra commuting with screenings \eqref{scr} depends on one continuos parameter $b$ ($a=\sqrt{1+b^{2}}$). It is non-rational CFT with continuos spectrum and
central charge \eqref{centra-charge}. It is natural to identify it with the $W-$algebra of the coset
\begin{equation}\label{coset}
   \frac{O(n+2)_{-K}}{O(n+1)_{-K}},\quad\text{where}\quad b^{2}=K-n.
\end{equation}
The weak coupling in our perturbed CFT corresponds to $K$ close to $n$ and the strong coupling to $K\gg n$. The strong coupling asymptotic can be described by the SM-representation of our QFTs. The conformal limit of these QFTs \eqref{aan} gives us the SM - representation for the coset \eqref{coset}.

For positive and integer $-K=k$ the coset CFT (\ref{coset}) is the rational CFT with discrete spectrum. For large $k$ the low part of its spectrum  is 
\begin{equation}\label{spc1}
  \Delta\left(j_{1},\dots,j_{r}|i_{11},\dots,i_{r}\right)=\sum_{m=1}^{r_{n}}\frac{j_{m}(j_{m}+n-2m+2)}{2(k+n)}-
  \sum_{m=1}^{r_{n-1}}\frac{i_{m}\left(i_{m}+n-2m+1\right)  }{2(k+n-1)},
\end{equation}
where $r_{n}$ is the rank of $O(n+2)$ and  $j_{m}\geq j_{m+1},i_{m}\geq i_{m+1},j_{m}\geq i_{m}$.

It was shown in \cite{Altschuler:1988mg} that $\frac{O(n+2)_{k}}{O(n+1)_{k}}$ CFT is equivalent to $\frac{O(k)_{n+1}O(k)_{1}}{O(k)_{n+2}}$ coset models (the example of level-rank duality). The $W-$ algebras, its representations and minimal models corresponding to the $\frac{O(k)_{n+1}O(k)_{1}}{O(k)_{n+2}}$ cosets, where constructed and studied in \cite{Lukyanov:1989gg,Lukyanov:1990tf}. We call them $W(O(k))$-algebras. The minimal models of these algebras are specified by the level $n+1$ and integer $\mathrm{p}=n+k-1$.  The spectrum of dimensions of primary fields $\Phi\left(\Omega|\Omega^{\prime}\right)$  in these models  is characterized by two highest weights of the Lie algebra $O(k)$, $\Omega$ and $\Omega^{\prime}$, satisfying the conditions  $-\Omega\cdot e_{0}\leq n+2$, $-\Omega^{\prime}\cdot e_{0}\leq n+1$ and is
\begin{equation}\label{spc2}
  \Delta\left(\Omega|\Omega^{\prime}\right)=\frac{(\mathrm{p}\Omega-(\mathrm{p+1})\Omega^{\prime})^{2}-\frac{1}{2}k\left(k-1\right)\left(k-2\right)}{2(\mathrm{p+1})\mathrm{p}}.
\end{equation}
It is easy to check that the low lying spectrum \eqref{spc1} coincides with \eqref{spc2}). 

We see that  the parameter $b^{2}=-k-n$, of  $W-$ algebra studied in section \ref{CFT} is now imaginary ($a^{2}=-k-n+1$). It means that our $W-$algebra can be considered as ``horizontal'' $W-$ algebra of  $W-$ algebra $W(O(k))$, i.e. in our case  the rank changes and level is fixed. 

The minimal models of $\frac{O(k)_{n+1}O(k)_{1}}{O(k)_{n+2}}$  $W-$algebras can be used for the analysis of $O(n+2)$ SMs. It was shown in \cite{Fateev:1990bf,Fateev:1991bv} for $n=1$ and in \cite{Fendley:1999gb} for all $n$ that these SMs can be derived as the limit $k\rightarrow\infty$ of these minimal models perturbed by the field $\Phi\left(0|\omega_{1}\right)$, where fundamental weight $\omega_{1}$ corresponds to the vector representation. 
\item The appearance of minimal models of $\frac{O(k)_{n+1}O(k)_{1}}{O(k)_{n+2}}$ with fixed level depending on $n$, which are established by our $W-$ algebra is rather natural. The analysis of minisuperspace equations for $T-$ dual metric (to deformed $O(n+2)-$SM metric) shows that in the conformal limit they have the discrete spectrum, which coincides with \eqref{spc1}) up to $\frac{1}{\mathrm{p}}$. It means that corresponding CFT flows from UV critical point to IR critical point, which is described by minimal models \eqref{spc2}), where
$\mathrm{p}=n+k-1$. When we have such RG flow in the minimal models should be the receiving $\Phi_{R}$ operator for this flow. The perturbation with this IR relevant field $\Phi_{R}$ should be integrable. The analysis of integrable IR perturbations shows that only possible such field is $\Phi(2\omega_{1}|0)$ with dimension $1+\frac{n}{\mathrm{p}+1}$. With this field we can develop the IR perturbation theory adding to the action of CFT the perturbation $\lambda\Phi(2\omega_{1}|0)$. The exact relation between the parameter $\lambda$ and the mass $M$ can be derived by BA-method \cite{Fateev:1993av,Zamolodchikov:1995xk} and is
\begin{equation*}
   \left(\pi\lambda\right)^{2}=\frac{\mathrm{p}^{2}(\mathrm{p}-n)^{2}\Gamma\left(1+\frac{n}{\mathrm{p}+1}\right)\Gamma\left(1+\frac{3n}{\mathrm{p}+1}\right)}
   {3(\mathrm{p}+n)^{2}(\mathrm{p}+2n)^{2}\Gamma\left(1-\frac{n}{\mathrm{p}+1}\right)\Gamma\left(1-\frac{3n}{\mathrm{p}+1}\right)}
   \left(\frac{2(\mathrm{p}+1)}{nM}\right)^{\frac{4n}{\mathrm{p}+1}}.
\end{equation*}
With this relations we can calculate the IR corrections to observables. In particular the first correction to the central charge of our QFT at the circle of length $R$ will be
\begin{equation*}
  c=c_{\textrm{CFT}}+3\left(\pi\lambda\right)^{2}\frac{\Gamma^{2}\left(1+\frac{n}{\mathrm{p}+1}\right)\Gamma\left(1-\frac{2n}{\mathrm{p}+1}\right)}
  {\Gamma^{2}\left(1-\frac{n}{\mathrm{p}+1}\right)\Gamma\left(1+\frac{2n}{\mathrm{p}+1}\right)}\left(\frac{2\pi}{R}\right)^{\frac{4n}{\mathrm{p}+1}}.
\end{equation*}
This exact result can be compared with numerical analysis of non-linear integral equation following from scattering theory data.
\item The integrable perturbations to the CFT \eqref{perturbation} considered in this paper are not unique. We can add to the action \eqref{ac2} also the terms with integrals with
fields $\mu_{1}e^{b\varphi_{n}}$ or $\mu_{2}e^{2b\varphi_{n}}$. At $b\ll1$ we can transform these CFTs to the form convenient for perturbation theory in $b$ as it was done in section \ref{dual-L}. At $b\gg1$, these CFTs will have the dual representation with SM UV action \eqref{aan} together with the potential terms $M^{2}U_{i}(\chi_{1}/b,\dots\chi_{k}/b)$, $i=1,2$. This theory can be studied by perturbation theory in $\frac{1}{b}$. The explicit form of functions $U_{i}$ and the factorized scattering theory for these QFTs we describe in other publication.
\end{enumerate}
\section*{Acknowledgments}
The work of A.L. is  supported by Laboratory of Mirror Symmetry NRU HSE, RF Government grant, ag. N 14.641.31.0001.
\Appendix
\section{Minisuperspace differential equation}\label{mss-diff}
It is convenient to define the function
\begin{equation*} 
   a(S)=\frac{1}{2}+iS,\quad a(-S)  =1-a(S),\quad a(S)a(-S)=\frac{1}{4}+S^{2}.
\end{equation*}
Then equation \eqref{HT} for the function $R(X,Y)\overset{\text{def}}{=}R_{P,Q}(x,y)$, where $X=e^{x}$ and $Y=e^{y}$ can be written in the form
\begin{multline*}
  0=\left(a_{P}^{2}X^{2}+a_{Q}^{2}\left(  2+X)Y\right)\right)R+Y\left(  (2+X)Y+a_{Q}\left(X+2Y+YX\right)\right)R_{Y}+\\
    +Y^{2}\left((2+X)Y+X\right)R_{YY}+X^{2}((Y+2a_{P}(X+1))R_{X}+X\left(X+1\right)R_{XX}.
\end{multline*}
This equation does not admit the separation of variables but it is useful to represent the solution in the form convenient for the analysis outside of ``Weyl chambre''. We will search the solution in the form
\begin{equation*}
  R(X,Y)=F(Y/X,Y)_{2}F_{1}(a_{P},a_{P},2a_{P},-X)-\frac{1}{a_{-P}}G(Y/X,Y)_{2}F_{1}(a_{P}-1,a_{P},2a_{P},-X),
\end{equation*}
where $_{2}F_{1}(a,b,c,X)$ is the hypergeometric function. To write down the equations for the functions $F(Z,Y)$ and $G(Z,Y)$ we denote $V=1+2Z+Y$.  Then  we have
\begin{multline}\label{yr1}
  0=\left(2Z+Y\right)  (a_{Q}^{2}F-G)+\left(Y+Z(1+V+2a_{Q}V\right))F_{Z}+(1+V+2a_{Q}V)YF_{Y}+\\+
   Z(Y+Z+ZV)F_{ZZ}-Z(Y+Z)G_{Z}+V(Y^{2}F_{YY}+2ZYF_{YZ})
\end{multline}
and
\begin{multline}\label{yr2}
  0=\left(2a_{Q}+\left(1+a_{Q}\right)^{2}(Z+Y)\right)G+Y\left(3V-1+2a_{Q}V\right)G_{Y}+2a_{P}a_{-P}F_{Z}+\\+
   \left(Y+Z(1+V+2a_{Q}V\right)  )G_{Z}+Y^{2}VG_{YY}+Z\left(2YVG_{YZ}+(1+Z)(2Z+Y)G_{ZZ}\right).
\end{multline}
We note that $Z=\frac{Y}{X}=e^{-x+y}\ll1$ is in the ``Weyl chambre''. We write our functions in this region as
\begin{equation*}
   F\left(Z,Y\right)=\sum_{n=0}^{\infty}F_{n}(Y)Z^{n},\quad G\left(Z,Y\right)=Z\sum_{n=0}^{\infty}G_{n}(Y)Z^{n}.
\end{equation*}
Then the solution to \eqref{yr1} and \eqref{yr2} can be expressed through two functions%
\begin{equation}\label{f0}
  F_{0}(Y)=_{4}F_{3}(a_{Q},a_{Q},a_{Q},a_{Q}|2a_{Q},a_{Q}+a_{P},a_{Q}+a_{-P}|-Y),
\end{equation}
and
\begin{multline}\label{g0}
  G_{0}(Y)=_{4}F_{3}(a_{Q},a_{Q},a_{Q}+1,a_{Q}+1|2a_{Q},a_{Q}+a_{P}+1,a_{Q}+a_{-P}+1|-Y)a_{Q}^{2}a_{P}a_{-P}\times\\\times
  \left((a_{Q}+a_{P}+1)(a_{Q}+a_{-P}+1)\right)^{-1}\,_{3}F_{2}(1,1,1|a_{Q}+a_{P}+1,a_{Q}+a_{-P}+1|-Y),
\end{multline}
by the relations
\begin{multline}\label{Fn}
-(1+n)^{2}YF_{n+1}=\left(a_{Q}+n\right)(2n+(a_{Q}+n)Y)F_{n}+(Y+2(a_{Q}+n)(1+Y))F_{n}^{\prime}-\\-
  Y(2n+1)G_{n}-2nG_{n-1}+Y(2F_{n-1}^{\prime\prime}+\left(1+Y\right)F_{n}^{^{\prime\prime}}),
\end{multline}
and $(-(1+n)^{2}YG_{n+1}+2(1+n)a_{P}a_{-P}F_{n+1})$ is
\begin{multline}\label{Gn}
  (2(n+1)(n+a_{Q})+(n+1+a_{Q})^{2})G_{n}+Y(2+3Y+2(a_{Q}+n)(1+Y))G_{n}^{\prime}+\\+
  2(n+a_{Q})^{2}G_{n-1}+Y(2G_{n-1}^{\prime\prime}+\left(1+Y\right)G_{n}^{^{\prime\prime}})+2(2n+1+2b)G_{n-1}^{\prime}.
\end{multline}

The hypergeometric functions $_{m+1}F_{m}(a_{1},..,a_{m+1}|b_{1},..,b_{m}|-W)$ possess the integral representation, which defines their series expansion in $W$ for small $W$ and is useful for calculation their asymptotic for $W\gg1$
\begin{equation}\label{hf}
   _{m+1}F_{m}\left(-W\right)=\int_{C}ds\,W^{s}\frac{\Gamma\left(-s\right)\Gamma(a_{1}+s)\times\dots\times\Gamma(a_{m+1}+s)\Gamma(b_{1})\times
    \dots\times\Gamma(b_{m})}{2\pi i\Gamma(b_{1}+s)\times\dots\times\Gamma(b_{m}+s)\Gamma(a_{1})\times\dots\times\Gamma(a_{m+1})}.
\end{equation}
The contour $C$ for $\operatorname{Re}a_{i}>0$ goes from $-i\infty$ to $i\infty$ including point $s=0$.

The asymptotic of the function $\Omega(P,Q|X,Y)=X^{iP}Y^{iQ}R(P,Q|X,Y)$ in the region $X\gg1$, $Y\gg1$ can be derived using \eqref{hf}. Namely function depending on $X$
\begin{equation}
\begin{aligned}
  X^{iP}\,_{2}F_{1}(a_{P},a_{P},2a_{P},-X)  &  =\frac{1}{X^{1/2}}\frac{\Gamma(1+2iP)}{\Gamma^{2}(\frac{1}{2}+iP)^{2}}(\log(X)+c_{1}+O(1/X\log(X))\\
  X^{iP}\,_{2}F_{1}(a_{P},a_{P},2a_{P},-X)  &  =\frac{X^{1/2}}{a_{-P}}\frac{\Gamma(1+2iP)}{\Gamma^{2}(\frac{1}{2}+iP)^{2}}(1+O(1/X\log(X))),
\end{aligned}
\end{equation}
and functions depending on $Y$
\begin{align*}
  Y^{iQ}F_{0}(P,Q,Y) &  =\frac{H\left(  P,Q\right)  }{Y^{1/2}}\left(\log^{3}(Y)+c_{2}\log^{2}(Y)+c_{3}\log(Y)+c_{4}+O(1/Y\log(Y)^{3})\right),\\
  Y^{iQ}G_{0}(P,Q,Y) &  =\frac{H\left(  P,Q\right)  }{Y^{3/2}}\left(\log^{3}(Y)+c_{5}\log^{2}(Y)+c_{6}\log(Y)+c_{7}+O(1/Y\log(Y)^{3})\right),
\end{align*}
here $c_{i}$ are some constant and
\begin{equation*}
  H=\frac{\Gamma(2a_{Q})\Gamma(a_{Q}+a_{P})\Gamma(a_{Q}+a_{-P})}{\Gamma(a_{Q})^{4}\Gamma(a_{P})\Gamma(a_{-P})}=
  \frac{\Gamma(1+2iQ)\Gamma(1+iQ+iP)\Gamma(1+iQ-iP)}{\Gamma(\frac{1}{2}+iQ)^{4}\Gamma(\frac{1}{2}+iP)\Gamma(\frac{1}{2}-iP)}.
\end{equation*}
The functions $F_{n}$, $G_{n}$ with $n>0$ have the asymptotic
\begin{equation*}
  F_{n}=\frac{f_{n}H}{Y^{1/2}}\left(\log(Y)+c_{n}+O(\log^{3}\left(Y\right)/Y)\right),\quad
  G_{n}=\frac{g_{n}H}{Y^{3/2}}\left(\log(Y)+c_{n}^{\prime}+O(\log^{3}\left(Y\right)/Y)\right)
\end{equation*}
where the constants $f_{n}$ and $g_{n}$ can be easily derived from \eqref{Fn}, \eqref{Gn}.
We see that the function $\Omega_{P,Q}(x,y)$ contains logarithmic terms and the dangerous for normalizability term, coming from the function $_{2}F_{1}(a_{P},a_{P},2a_{P},-X)$. 
However in the sum of the functions $\Omega_{P_{s},Q_{s}}(x,y)$ ($s\in w_{2}$) over the Weyl group $w_{2}$ with coefficients $\mathrm{A}^{(\textrm{cl})}(P_{s},Q_{s})$ (and only with these coefficients) all logarithmic and dangerous terms dissapear and  Weyl invariant function $\Psi_{P,Q}(x,y)$
\begin{equation}\label{SOLUT}%
  \Psi_{P,Q}(x,y)=\sum_{s\in w2}\mathrm{A}^{(cl)}(P_{s},Q_{s})\Omega_{P_{s},Q_{s}}(x,y),
\end{equation}
where
\begin{equation*}
  \mathrm{A}^{(\textrm{cl})}(P,Q)=\frac{\Gamma(-2iQ)\Gamma(-2iP)\Gamma(-iQ-iP)\Gamma(iP-iQ)\Gamma(\frac{1}{2}+iP)\Gamma(\frac{1}{2}+iQ)^{2}}
  {\Gamma(\frac{1}{2}-iQ)^{2}\Gamma(\frac{1}{2}-iP)}
\end{equation*}
gives us unique solution which has a regular expansion for $X\gg1$, $Y\gg1$
\begin{multline*}
  (XY)^{1/2}\Psi_{P,Q}(x,y)=1-\frac{p^{2}q^{2}}{Y}+\frac{(p^{2}+q^{2}-p^{2}q^{2})}{X}+\frac{p^{2}q^{2}(4+2p^{2}+2q^{2}-p^{2}q^{2})}{4YX}+\\
  +\frac{p^{2}(2+p^{2})q^{2}(2+q^{2})}{16Y^{2}}+\frac{((p^{2}+q^{2})\left(4-2(p^{2}+q^{2})-3p^{2}q^{2}\right)  -4p^{2}q^{2}+\frac{p^{4}q^{4}}{2})}{8X^{2}}+\dots
\end{multline*}
where $p^{2}=P^{2}+\frac{1}{4}$ and $q^{2}=Q^{2}+\frac{1}{4}$.
\section{Integrals of Motion}\label{IM}
In this appendix we present explicit expressions for the density of first non-trivial local Integral of Motion in our theory \eqref{ann},\eqref{acd} which is dual to the deformed $O(n+2)$ sigma-model. We define the density $G(x)$ through the relation
\begin{equation}
   \mathbf{I}_{3}=\int G(x)dx
\end{equation}
We denote $Q_{i}=\partial\phi_{i}$, $P_{i}=\partial\varphi_{i}$ and $a^{2}=1+b^{2}$. First examples of  the density $G(x)$ have the form
\begin{enumerate}
\item  $n=1$
\begin{equation}
  G=AQ^{4}+BP^{4}+GQ^{2}P^{2}+RQ^{\prime2}+TP^{\prime2},
\end{equation}
where
\begin{equation*}
   A=(1+b^{2})(1+3b^{2}),\,B=b^{2}(2+3b^{2}),\,G=6b^{2}\left(1+b^{2}\right),\quad R=1+7b^{2}+8b^{2},\quad T=2+9b^{2}+8b^{4}.
\end{equation*}
\item $n=2$
\begin{equation}
  G=A(Q_{1}^{4}+Q_{2}^{4})+BP^{4}+6a^{2}b^{2}((Q_{1}^{2}Q_{2}^{2}+(Q_{1}^{2}+Q_{2}^{2})P^{2})+MP^{\prime}(Q_{1}^{2}-Q_{2}^{2})+
  R(Q_{1}^{^{\prime}2}+Q_{2}^{\prime2})+SP^{\prime2}
\end{equation}
where
\[
A=B=3a^{2}b^{2},\,M=-6a^{2}b,\,R=6a^{2}b^{2},\,S=3a^{2}(1+2b^{2}).
\]
\item $n=4$
\begin{multline}
   G=A(Q_{1}^{4}+Q_{2}^{4}+Q_{3}^{4})+B(P_{1}^{4}+P_{2}^{4})+6a^{2}b^{2}(Q_{1}^{2}Q_{2}^{2}+Q_{1}^{2}Q_{3}^{2}+Q_{2}^{2}Q_{3}^{2})+\\
   +6a^{2}b^{2}((Q_{1}^{2}+Q_{2}^{2}+Q_{3}^{2})(P_{1}^{2}+P_{2}^{2})+P_{1}^{2}P_{2}^{2})+\\
   +M\left(  P_{1}^{\prime}(Q_{1}^{2}-Q_{2}^{2}-Q_{3}^{2}-P_{2}^{2})-P_{2}^{\prime}(-Q_{1}^{2}-Q_{2}^{2}+Q_{3}^{2}-P_{1}^{2})\right)+\\
   R(Q_{1}^{\prime2}+Q_{2}^{\prime2}+Q_{3}^{\prime2})+S\left(P_{1}^{\prime2}+P_{2}^{\prime2}\right)+WP_{1}^{\prime}P_{2}^{\prime}
\end{multline}
where
\begin{multline*}
  A=a^{2}\left(2a^{2}+b^{2}\right),\,B=b^{2}\left(a^{2}+2b^{2}\right),\,M=6a^{2}b,\,R=4-2a^{2}+10a^{2}b^{2},\\
  S=a^{2}-2b^{2}+10a^{2}b^{2},\,W=-6a^{2}.
\end{multline*}
\end{enumerate}
We note that in the limit $b\rightarrow\infty$ the Integral $\mathbf{I}_{3}$ enjoys $O(n+1)$ symmetry and coincides with the one introduced in \cite{Lukyanov:2003rt}.

\bibliographystyle{MyStyle}
\bibliography{MyBib}

\end{document}